\documentclass{aa}

\usepackage{graphicx}
\usepackage{soul}
\usepackage{multirow}
\usepackage{float}
\usepackage{txfonts}
\usepackage{placeins}
\usepackage{verbatim}
\usepackage{appendix}
\usepackage{lipsum}
\usepackage{dblfloatfix}
\usepackage{caption}

\usepackage{afterpage}

\usepackage[
  breaklinks = true,
  colorlinks = true,
  urlcolor   = blue,
  citecolor  = blue,
  linkcolor  = blue,
]{hyperref}
\usepackage{xcolor}
\usepackage[labelfont=bf]{caption}

\begin{document} 

    \title{The magnetic sensitivity of the Ca~{\sc{ii}} resonance and subordinate lines in the solar atmosphere}

    \titlerunning{Ca~{\sc{ii}} resonance and subordinate lines} \authorrunning{I\~nigo Juanikorena et al.}
    
   \author{I. Juanikorena Berasategi \inst{1, 2}, E. Alsina Ballester \inst{1, 2}, T. del Pino Alemán \inst{1, 2}, and J. Trujillo Bueno \inst{1, 2, 3, }\thanks{{Affiliate Scientist of the National Center for Atmospheric Research, Boulder, CO, USA.}}}

   \institute{Instituto de Astrofísica de Canarias, E-38205 La Laguna, Tenerife, Spain
         \and
             Departamento de Astrofísica, Universidad de La Laguna, E-38206 La Laguna, Tenerife, Spain
        \and
            Consejo Superior de Investigaciones Cient\'ificas, Spain
            }

   \date{Received July 10, 2025; accepted }

  \abstract{}
   {The polarization of the Ca~{\sc{ii}} resonant doublet (H and K lines) and subordinate infrared triplet lines are valuable observables for diagnosing the magnetism of the solar chromosphere. It is thus necessary to understand in detail the physical mechanisms that play a role in producing their Stokes profiles in the presence of magnetic fields.}
   {We use the spectral synthesis module of the HanleRT-TIC code to study the impact 
   of anisotropic radiation pumping with partial frequency redistribution (PRD) and J-state interference (JSI), considering a plane-parallel semi-empirical static solar atmospheric model. We also study the sensitivity of the lines to magnetic fields of various strengths and orientations accounting for the joint action of the Hanle and Zeeman effects.}
   {Taking PRD into account is required to suitably model the polarization in the core regions of the resonant lines, whereas JSI plays a crucial role in their far wings. We confirm that the metastable lower levels of the subordinate lines also contribute to the scattering polarization of the K line. In the presence of horizontal magnetic fields, we find that the resonant lines are sensitive to a wide range of field strengths (sub-gauss to tens of gauss), whereas the scattering polarization of the infrared triplet lines are mainly sensitive to milligauss field strengths. At a near-limb line of sight (LOS) with $\mu = 0.1$, the Hanle effect modifies the scattering polarization via a depolarization and a rotation in the plane of linear polarization. At disk center, horizontal fields in the 1D semi-empirical model give rise to linear polarization signals; in the K line, this is governed by the Hanle effect in the sub-gauss to few tens of gauss range and by the Zeeman effect in stronger fields. For vertical magnetic fields, the Hanle effect does not operate, but the linear polarization wings of the resonance lines are sensitive to magneto-optical effects. Finally, we find that the atomic level polarization exerts an influence on the outer circular polarization lobes of the resonant lines and that the weak field approximation tends to overestimate the LOS magnetic field component if this frequency range is considered.}
   {}
   

   \maketitle

\section{Introduction}
The spectral lines of the ultraviolet (UV) resonant doublet and the infrared (IR) subordinate triplet of Ca~{\sc{ii}} are of great interest in the study of the solar chromosphere. The transitions that give rise to these five spectral lines share the same upper atomic term; the H and K lines of the UV doublet are produced by transitions from this term to the Ca~{\sc{ii}} ground level, whereas the IR triplet lines are due to transitions to the lowest-energy metastable term. The H and K resonant lines originate in the upper chromosphere (e.g., \citealt{Ca_II_K_formation_height}), whereas the lines of the IR triplet encode information on the middle chromosphere (see Figure 3 in \citealt{MS_2014}). These spectral lines exhibit scattering polarization (i.e., the polarization produced by the scattering of anisotropic radiation) and their polarization profiles are sensitive to magnetic fields via the Zeeman and Hanle effects, therefore being of significant diagnostic potential for the investigation of the chromospheric magnetism (e.g., the review by \citealt{Review_2022_TB_Tanausu}). \\

Observational data of the infrared $8542 \: \AA$ line have been extensively used to study the magnetism of the solar atmosphere (e.g., \citealt{Type_II_spicules_Ca_II_8542, Longitudinal_B_QS_Ca_II_8542,  Chromospheric_grains_Ca_II_K}). The CRisp Imaging SpectroPolarimeter [CRISP, \cite{CRISP}] at the 1-m Swedish Solar Telescope [SST, \cite{SST}] and, more recently, the GREGOR Infrared Spectrograph [GRIS, \cite{GRIS, GRIS_Gregor_IFU}, \cite{GRIS_3}] at the GREGOR telescope \citep{GREGOR} enable spectropolarimetric observations of this subordinate line. In recent years, new technical capabilities have been developed that will provide a large volume of spectropolarimetric data and allow further investigations using the resonance and subordinate lines of Ca~{\sc{ii}}. The Daniel K. Inouye Solar Telescope \citep{DKIST}, currently the largest ground-based solar observation facility, can observe the intensity and polarization of all five lines considered in this work simultaneously with its Visible Spectropolarimeter [ViSP, \cite{VISP}]. Furthermore, the recently launched SUNRISE III balloon-borne mission \citep{Sunrise_III} has acquired a large amount of unprecedented data. The Sunrise Ultra-violet Spectropolarimeter and Imager [SUSI, \cite{SUSI}] instrument was designed for the spectral range between $300 \: \mathrm{nm}$ and $410 \: \mathrm{nm}$ (including the resonance lines), whereas the SUNRISE Chromospheric Infrared spectroPolarimeter [SCIP, \cite{SCIP}] instrument was specifically developed for a few near-IR lines including Ca~{\sc{ii}} $8542 \: \AA$. Considering all this, we expect a great abundance of scientific work in the following years related to these lines of Ca~{\sc{ii}}.\\

A correct interpretation of spectropolarimetric observations in the Ca~{\sc{ii}} lines requires a deep understanding of the physical processes that produce their intensity and polarization. Such knowledge can be gained through numerical investigations based on spectral synthesis, which make use of radiative transfer (RT) calculations. The chromospheric spectral lines under consideration form in atmospheric regions where radiation processes dominate over collisions, invalidating the local thermodynamic equilibrium (LTE) assumption. Under non-LTE conditions, the source function is not described by the Planck function and we are required to solve numerically the coupled RT and Statistical Equilibrium (SE) equations applying radiative transfer codes.\\

The scattering of anisotropic radiation can give rise to linear polarization in spectral lines, which is sensitive to the magnetic field via the Hanle effect \citep{Landi_2004}. The anisotropy of the radiation field generates atomic level polarization (i.e., population imbalances and quantum coherence among magnetic sublevels). The polarized atomic levels may produce linear polarization signals either via the selective emission of light from the upper levels of each transition or via dichroism (i.e., the selective absorption of specific polarization states), due to the atomic level polarization of the lower levels \citep{TB_1997, JTB-1999, JTB-Nature-2002, Dichroism}. The modification of scattering polarization signals in the spectral lines of Ca~{\sc{ii}} via the Hanle effect increases their magnetic diagnostic value \citep{MSTB_2010, Stepan_TB_2016}. Moreover, the magnetic field can generate polarization signals, especially circular polarization, via the Zeeman effect.\\

Strong resonant chromospheric lines, such as the $4227 \: \AA$ line of Ca~{\sc{i}} and the UV doublets of either Mg~{\sc{ii}} or Ca~{\sc{ii}}, often require taking into account partial frequency redistribution (PRD) effects for a correct modeling of their Stokes spectra (e.g., \citealt{Mg_II_HK_Belluzzi, Mg_II_Ernest, HanleRT, Alsina_2017, HanleRT_Mg_II}). The PRD treatment implies accounting for the frequency correlations between the incoming and outgoing radiation in scattering processes, which makes numerical calculations much more computationally expensive. Additionally, for strong resonant multiplets, the quantum-mechanical interference between the upper atomic J levels (i.e., J-state interference) can substantially modify the polarization profiles that are produced in the wings of the spectral lines due to PRD. J-state interference leaves clear signatures in the spectral range between the lines of the multiplet \citep{Mg_II_HK_Belluzzi, HanleRT, Mg_II_Ernest_2}. \\

In the present work, we investigate the impact of including PRD effects and J-state interference in the UV doublet and the IR triplet of Ca~{\sc{ii}}, while also analyzing their magnetic sensitivity. We employed the publicly available radiative transfer code HanleRT-TIC\footnote{\url{https://gitlab.com/TdPA/hanlert-tic}} \citep{HanleRT, HanleRT_Mg_II, HanleRT_TIC} to generate spectral syntheses of the Ca~{\sc{ii}} resonance and subordinate lines including all the aforementioned physics. We aim to clarify which are the key physical mechanisms that control the polarization in these lines. In section \ref{Section_Formulation}, we present the details of the radiative transfer calculations carried out for this investigation. Section \ref{Section_Unmag} shows how the frequency redistribution treatment of each line and the atomic model selected for Ca~{\sc{ii}} affect the emergent intensity and linear polarization profiles of the lines of interest, while section \ref{Section_Mag} analyzes their magnetic sensitivity when imposing magnetic fields with strengths ranging from a few milligauss to a few hundreds of gauss. Finally, in section \ref{Section_circular_pol}, we present the circular polarization profiles produced by the Zeeman effect.\\

\section{Modeling approach}\label{Section_Formulation}
In this investigation, we conduct a theoretical analysis of the physics that must be taken into account for accurately modeling the Stokes profiles of the Ca~{\sc{ii}} lines, and to study their magnetic sensitivity. For this goal, we carried out a series of spectral syntheses using HanleRT-TIC \citep{HanleRT, HanleRT_TIC}, a non-LTE radiative transfer code for the synthesis and inversion of Stokes profiles in one-dimensional (1D) atmospheric models. This radiative transfer code accounts for atomic level polarization, scattering polarization, J-state interference, PRD effects, and the impact of the magnetic field through the joint action of the Hanle and Zeeman effects. In this section we describe the modeling approach selected for the syntheses.

\subsection{The atmospheric model}
The synthetic profiles presented throughout this work were carried out using the atmospheric model C presented in \citet{FAL_C}, hereafter FAL-C. This one-dimensional (1D) plane-parallel semi-empirical model is representative of an average region of the quiet solar atmosphere. The line-of-sight (LOS) direction of the emergent radiation is given by two angles, the inclination with respect to the local vertical $\theta$ (with its cosine $\mu = \cos\theta$) and its azimuth $\chi$. Throughout this work, we set this azimuth to zero. The magnetic field, when present, is likewise given by its inclination $\theta_B$ and azimuth $\chi_B$, which is taken to be equal to $\chi$ in all cases, so the magnetic field vector is coplanar with the local vertical and the LOS, pointing towards the observer. This geometric setup is displayed in Figure 1 of \citet{Alsina_2017}. In Appendix \ref{Appendix_FAL_P}, we provide some equivalent plots to those presented in the results sections, which were instead computed considering the atmospheric model P of \citet{FAL_C} (FAL-P), a semi-empiric atmospheric model representative of a solar plage.

\subsection{The atomic system}
In order to model the intensity and polarization in the spectral lines considered in this work (H, K, $8498 \: \AA$, $8542 \: \AA$, $8662 \: \AA$), a suitable atomic representation requires only the five following levels (\citealt{MSTB_2010}, and more references therein): the ground level ($4 s\, {}^2\mathrm{S}_{1/2}$), two upper levels in the $4p\,{}^2\mathrm{P}$ term ($4p\,{}^2\mathrm{P}^{\mathrm{o}}_{1/2}$ and $4p\,{}^2\mathrm{P}^{\mathrm{o}}_{3/2}$), and two metastable levels in the $3d\, {}^2\mathrm{D}$ term ($3d\, {}^2\mathrm{D}_{3/2}$ and $3d\, {}^2\mathrm{D}_{5/2}$). The H and K lines (i.e., the UV doublet) arise from the transitions between the upper $4p\,{}^2\mathrm{P}$ and the ground $4 s\, {}^2\mathrm{S}$ terms. The lines of the IR triplet arise from transitions between the upper $4p\,{}^2\mathrm{P}$ and the metastable $3d\, {}^2\mathrm{D}$ terms. Further details about the treatment of the elastic and inelastic collisions are given in Appendix \ref{Appendix_collisions}. The atomic model and the transitions that give rise to the spectral lines are presented in the Grotrian diagram in Figure 1 of \citet{MSTB_2010}.\\

The calculations with HanleRT-TIC can consider multi-level or multi-term treatment of the atomic system. Whereas, in the former case, J levels belonging to the same term are treated as fully independent, in the latter case the quantum interference between such levels (J-state interference) is taken into account. To ensure consistency in the SE equations, the radiation spectrum must be flat over a frequency range at least as broad as the largest frequency separation for which quantum interference is considered. Disregarding the flat-spectrum condition can lead to non-physical results in the complete frequency redistribution (CRD) approximation. In the multi-level case, only interference between magnetic sublevels of each J level is accounted for, so the flat-spectrum contribution to the emission vector can consider different radiation fields for transitions between different levels. In contrast, the multi-term case includes interference between different J levels within the same term, so the spectrum must be flat across the whole multiplet, i.e., it must span the whole frequency range of the spectral lines involved. This can give rise to inaccuracies if such transitions are widely separated; thus, the multi-term treatment can be problematic for the IR triplet lines, which are separated by more than $160~\AA$. For a more detailed explanation of the underlying theory, see \cite{Flat_spectrum}.\\

Our atomic system corresponds to a multi-level atom comprised of five levels (5L) or a multi-term atom with three terms (3T). In section \ref{Subsection_Unmag_metastable}, we investigate the necessity of including the metastable levels in the atomic system for modeling the H and K lines. If we neglect the metastable levels (or the metastable term) in the calculation, the atomic system corresponds to a multi-level atom with three levels (3L) or a multi-term atom with two terms (2T).\\

\vspace{-0.3cm}
\subsection{The frequency redistribution treatments}
We also investigate the suitability of different treatments of the frequency redistribution for the synthesis of the Ca~{\sc{ii}} lines. The numerically simplest treatment we considered is that of CRD, in which the frequencies of the incoming and outgoing radiation in the scattering processes are treated as fully uncorrelated. This is equivalent to taking the incident radiation field to be spectrally flat within the relevant frequency interval. The treatment of PRD does account for the correlation between the incident and scattered radiation, which requires a more involved and numerically intensive calculation. In this work, the PRD calculations were carried out under the angle-averaged (AA) approximation \citep{Angle_averaged_RD, Belluzzi_TB_2014}; a comparison between such results and those of the full angle-dependent (AD) case will be presented in a forthcoming publication. The explicit form of branching ratios of the CRD and PRD contributions to the emissivity is presented in \citet{Flat_spectrum, Casini_2017b}. Finally, because PRD effects are particularly important for strong resonance lines, we also considered syntheses in which the PRD treatment is made for the H and K lines, whereas a CRD treatment is made for the lines of the IR triplet. Hereafter, we refer to this as the IRCRD treatment. When taking into account the effects of PRD for the H and K lines, we considered cross-redistribution (also referred to as Raman scattering) to the IR triplet.\\

\vspace{-0.3cm}
\subsection{The sensitivity to the magnetic field}
In the absence of magnetic fields, the linear polarization in the Ca~{\sc{ii}} K line arises as a result of the atomic alignment of its upper level. The linear polarization of the IR triplet lines, in addition to their upper levels, is sensitive to the atomic polarization of their lower, metastable levels, which give rise to selective absorption of polarization components (i.e., ``zero-field'' dichroism). Indeed, the scattering polarization of the $8662 \: \AA$ line, whose upper level has $J = 1/2$, is only due to zero-field dichroism \citep{Dichroism}. In the presence of magnetic fields, the atomic polarization of the upper and lower levels of such lines is modified by the Hanle effect, which leaves its signature on the linear polarization in the core region of these lines \citep{JTB-Review-2001, Landi_2004}. The Hanle effect operates when the Larmor frequency of the considered level is comparable to its mean lifetime. The efficiency of the Hanle effect in modifying (by dephasing and reducing) the atomic polarization of a given level is quantified by the ratio of the Larmor frequency to the mean lifetime of the level, i.e.,

\begin{equation} \label{Eq_Hanle_parameter}
   \quad \quad \quad \quad H_{u} = \frac{0.878 g_{u} B}{\sum_{\ell}A_{u\ell}} \quad \quad ; \quad \quad H_{\ell} = \frac{0.878 g_{\ell} B}{\sum_{u}B_{\ell u} J_{u \ell}} \quad ,
\end{equation}

\noindent where the subscripts $u$ and $\ell$ indicate the upper and lower level of a transition, respectively. $B$ is the strength of the magnetic field (in $\mathrm{G}$), and $g_{u}$ and $g_{\ell}$ are the Land\'e factors for the considered upper and lower levels, respectively. The sum over $\ell$ comprises all levels that are radiatively coupled to the considered upper level through spontaneous emission, with $A_{u \ell}$ being the corresponding Einstein coefficients (in $10^{7} \mathrm{s}^{-1}$ units) for spontaneous emission. The sum over $u$ goes over all the upper levels coupled to the considered lower level through radiative absorption, with $B_{\ell u}$ the corresponding Einstein coefficient for absorption and $J_{u\ell}$ the mean intensity at the wavelength of the transition (in $10^{7} \mathrm{s}^{-1}$ units for $B_{\ell u} J_{u \ell}$). \\ 

The Hanle critical field, being the magnetic field strength that characterizes the onset of the Hanle effect for the atomic level under consideration, can be found from Equation (\ref{Eq_Hanle_parameter}) by setting $H_u = 1$ or $H_\ell = 1$. For the upper level $4p\,{}^2\mathrm{P}^{\mathrm{o}}_{3/2}$, $B_{\mathrm{Hanle}} \approx 15 \: \mathrm{G}$ and, for the metastable levels, $B_{\mathrm{Hanle}}$ is $\sim \mathrm{mG}$. Thus, the upper level of the K line is sensitive to the upper-level Hanle effect for field values of several gauss, whereas the IR triplet lines can be modified by the lower-level Hanle effect and their atomic polarization is sensitive to sub-gauss fields.\\

\vspace{-0.4cm}
\section{The unmagnetized case} \label{Section_Unmag}
In this section, we investigate how the redistribution treatment (subsection \ref{Subsection_Unmag_RD_models}) and atomic model (subsection \ref{Subsection_Unmag_atomic_models}) affect the emergent profiles of the UV doublet and the IR triplet. In subsection \ref{Subsection_Unmag_metastable}, we examine the effect that excluding metastable levels in the atomic model has on the K line. No magnetic field is considered in this section and the results of our RT calculations are shown for a near-limb LOS with $\mu = 0.1$. The reference direction for positive Stokes $Q$ is the parallel to the nearest limb. Thus, for the considered geometry, there are no $U$ signals. \\

\vspace{-0.4cm}
\subsection{Redistribution treatment} \label{Subsection_Unmag_RD_models}

    \begin{figure*}[t]
        \centering
        \includegraphics[width=0.85\textwidth]{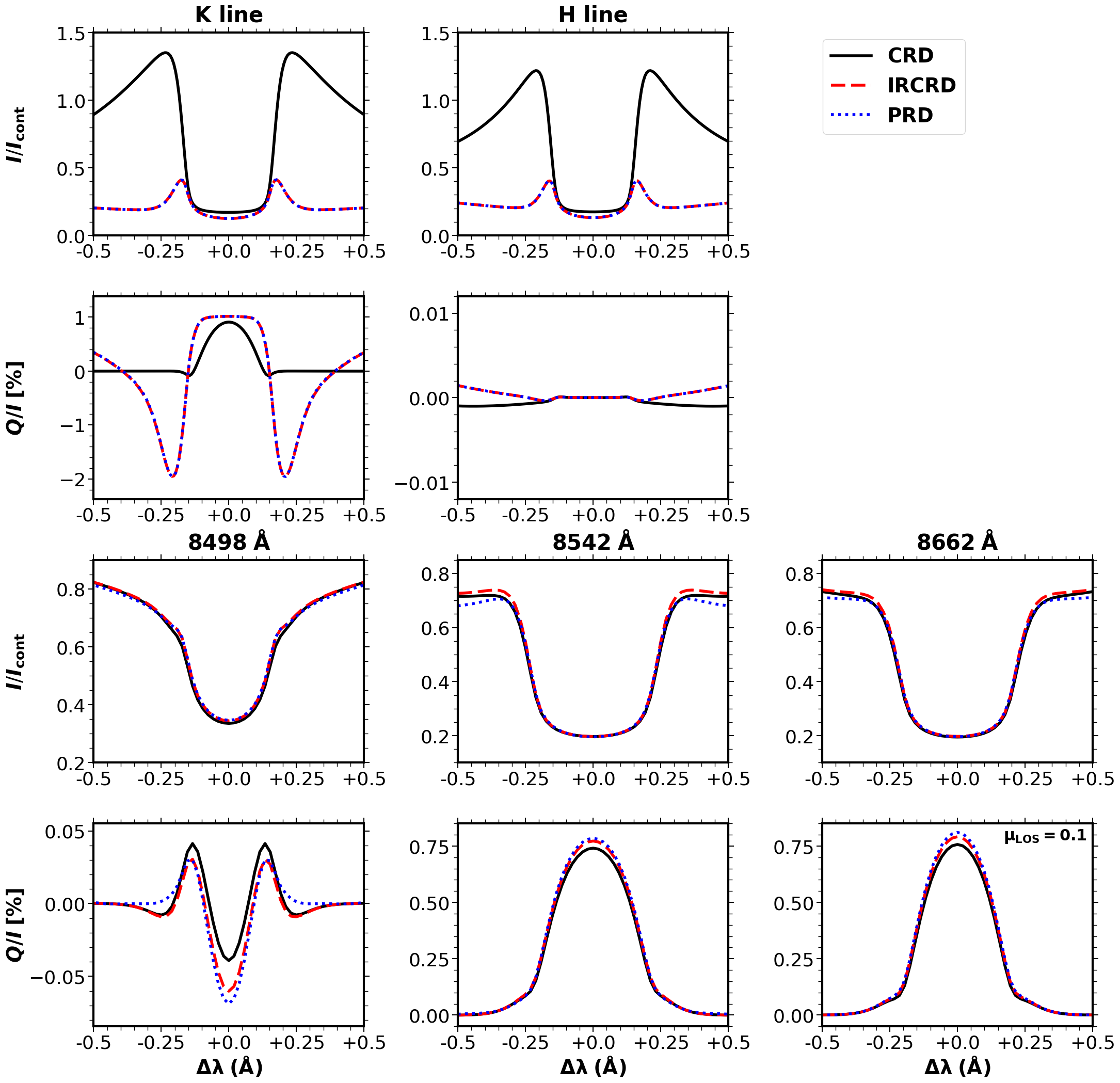}
        \caption{Stokes $I$, normalized to the continuum intensity $I_\mathrm{cont}$ (first and third rows), and fractional linear polarization $Q/I$ profiles (second and fourth rows), shown as a function of wavelength distance from the center of each line under consideration. The intensity profiles of the UV doublet and of the IR triplet lines are normalized to the continuum value at $\lambda = 3954.8 \: \AA$ and at $\lambda = 8500.5 \: \AA$, respectively. The profiles were computed in the semi-empirical model C of \cite{FAL_C} assuming a 5L atomic model. The emergent profiles are shown at a line of sight with $\mu = 0.1$. The various curves represent the results of different redistribution treatments: CRD (solid black), IRCRD (dashed red), and PRD (dotted blue). The reference direction for positive Stokes $Q$ is the parallel to the nearest limb.}
        \label{Fig_1_Unmag_all_lines_diff_RD_FALC}
    \end{figure*}

    \begin{figure*}[t]
        \centering
        \includegraphics[width=1.\textwidth]{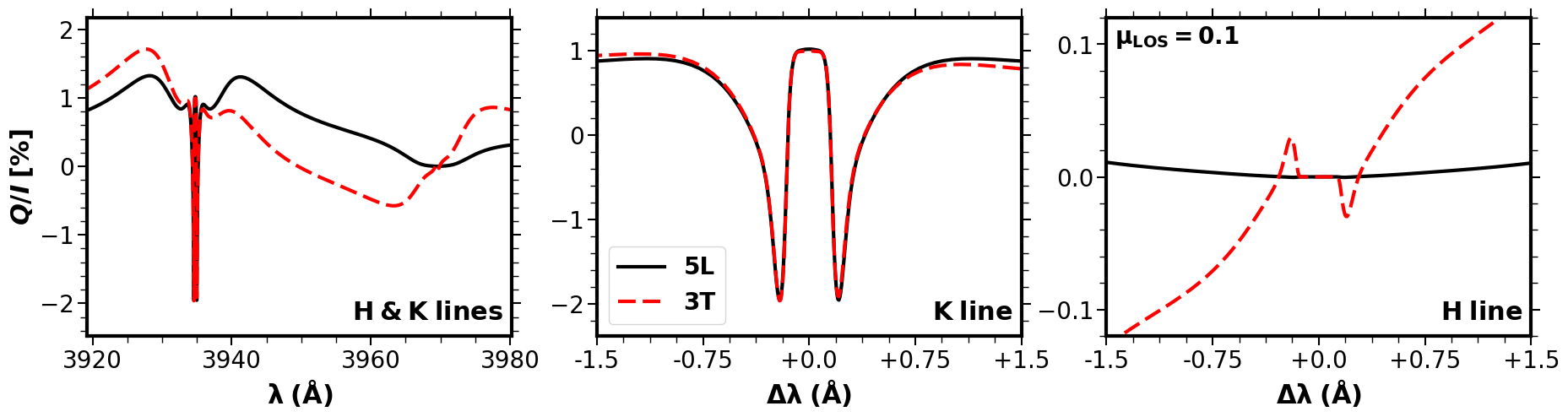}
        \captionof{figure}{Fractional linear polarization $Q/I$ profiles for the region around the H and K lines (left panel), and the core and near-wing region of the K line (central panel) and the H line (right panel). The considered redistribution treatment is IRCRD and the emergent profiles are shown for a LOS with $\mu=0.1$. The curves represent the emergent profiles for a 5L atomic model (which cannot account for J-state interference) in solid black, and a 3T model (which accounts for J-state interference) in dashed red. The reference direction for positive Stokes $Q$ is the parallel to the nearest limb.}
        \label{Fig_3_with_Hline_Unmag_HK_lines_diff_atoms_FALC}
    \end{figure*}

    \begin{figure*}[t]
        \centering
        \includegraphics[width=0.9\textwidth]{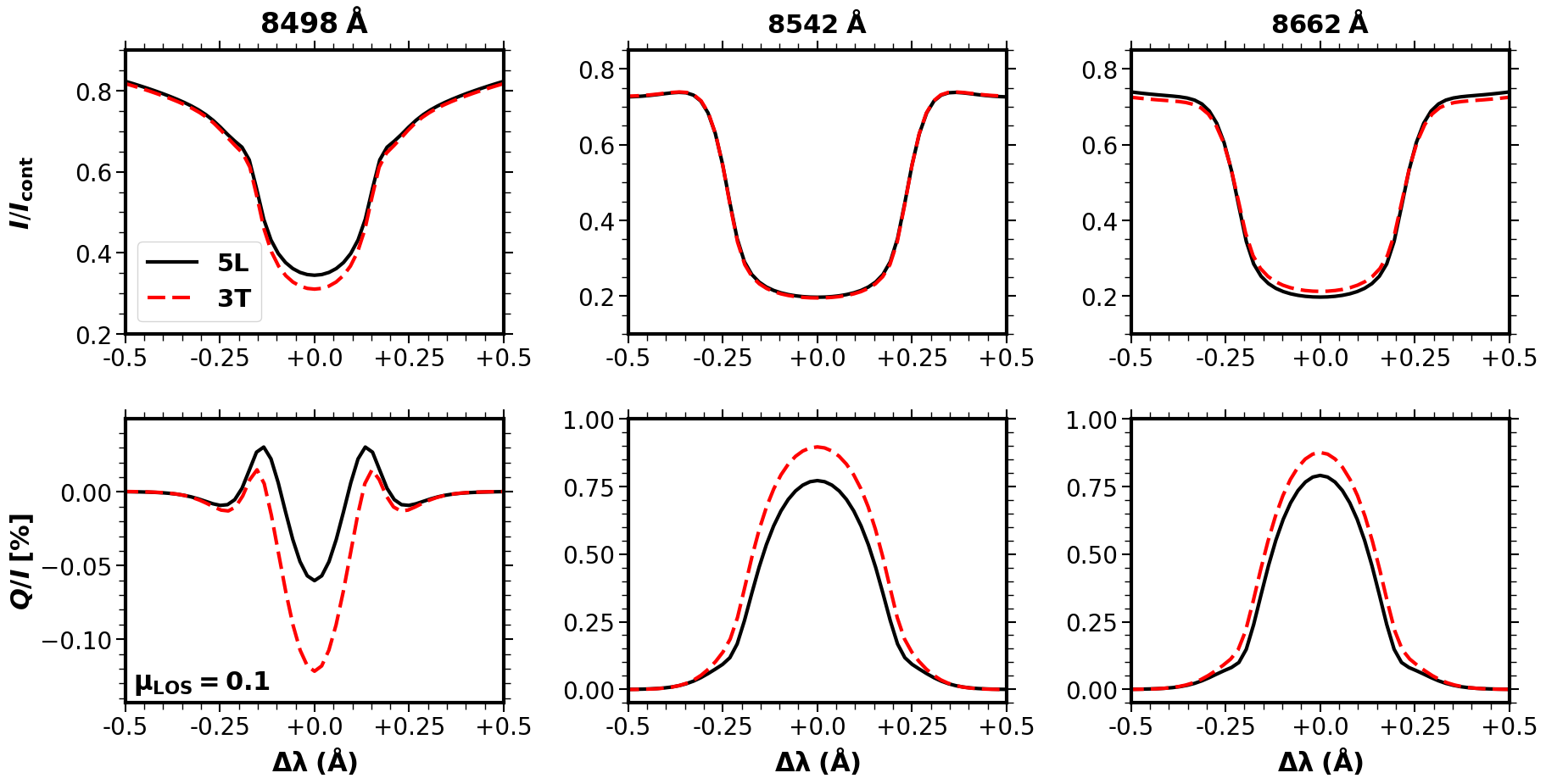}
        \captionof{figure}{Stokes $I$, normalized to the continuum intensity $I_\mathrm{cont}$ (upper panels), and fractional linear polarization $Q/I$ profiles (lower panels) for the IR triplet as a function of wavelength distance from the center of each line; from left to right, the $8498 \: \AA$, $8542 \: \AA$, and $8662 \: \AA$ lines. The emergent profiles are shown for a near the limb line of sight, with $\mu=0.1$. The curves represent emergent profiles for a 5L atomic model (which cannot account for J-state interference), in solid black, and a 3T model (which accounts for J-state interference), in dashed red. The reference direction for positive Stokes $Q$ is the parallel to the nearest limb.}
        \label{Fig_2_Unmag_IRT_lines_diff_atoms_FALC}
    \end{figure*}
    
    Figure \ref{Fig_1_Unmag_all_lines_diff_RD_FALC} shows the emergent continuum-normalized intensity ($I/I_\mathrm{cont}$) and fractional linear polarization ($Q/I$) as a function of wavelength, for the core region of the five lines under consideration. The figure compares profiles obtained with a five level atomic model considering three different redistribution treatments: CRD, IRCRD, and PRD.\\

    We find that the PRD treatment is necessary to correctly model the intensity and linear polarization of the H and K lines, especially beyond the line core. The $I/I_\mathrm{cont}$ and $Q/I$ profiles obtained in the CRD case differ from those found in the more general PRD case, even in the core region. We recall that the center of the H line cannot show linear scattering polarization signals because both its upper and lower levels have total angular momentum $J=1/2$, which prevents them from having atomic alignment.\\

    On the other hand, for the IR triplet, we find that the three redistribution treatments are in relatively good agreement. The $Q/I$ profile corresponding to the CRD case differs slightly from the other two, with maximum differences below $0.1 \%$ in the line center, which validates the results of \citet{Dichroism, MSTB_2010}. The IRCRD and the PRD results overlap in most of the frequency range, mainly in $Q/I$, although there are some differences in the intensity wings of the IR triplet lines. Differences in $I/I_\mathrm{cont}$ as large as $0.1$ are found at $\Delta \lambda \sim 1 \: \AA$, but all profiles converge by $\Delta \lambda \sim 2 \: \AA$. Unless otherwise noted, the IRCRD frequency treatment is considered in the following sections of this work.\\
    
    We also performed syntheses in unmagnetized, non-static atmospheres that include only vertical velocity gradients, or both horizontal and vertical velocity gradients, as described in Section 2 of \citet{Tanausu_AD}. We find that the qualitative conclusions presented in this section remain valid in the presence of such velocity gradients. The differences in the emergent profiles are small for the IR triplet lines and, when PRD is considered in the H and K lines (both IRCRD and PRD treatments), the resulting profiles overlap, similarly to that shown in Figure \ref{Fig_1_Unmag_all_lines_diff_RD_FALC}. A detailed analysis of the effects of velocity gradients in the magnetized case, while also including AD PRD, will be presented in a future work.\\
        
\subsection{Comparison of calculations with multi-level and multi-term atoms}\label{Subsection_Unmag_atomic_models}
    
    In order to analyze the impact of the atomic representation on the polarization profiles, we now show in Figures \ref{Fig_3_with_Hline_Unmag_HK_lines_diff_atoms_FALC} and \ref{Fig_2_Unmag_IRT_lines_diff_atoms_FALC} the resulting $Q/I$ profiles considering the multi-level (5L) and multi-term (3T) atomic systems for the UV lines and the IR triplet, respectively. In the former, only the fractional polarization profile is shown because $I$ overlaps perfectly for both cases. In the following, only the polarization profiles are shown when intensity profiles are unchanged when modifying a given parameter. \\

    An excellent agreement is found in the core and near wings of the K line, as shown in the central panel of Figure \ref{Fig_3_with_Hline_Unmag_HK_lines_diff_atoms_FALC}. However, there are clear differences in the wings for $\Delta \lambda > 0.75 \: \AA$. The largest discrepancy between the results from the two atomic representations is found in the frequency range between the H and K lines (see left panel of Figure \ref{Fig_3_with_Hline_Unmag_HK_lines_diff_atoms_FALC}); it can be attributed to J-state interference, which is taken into account in the multi-term, but not the multi-level case. Furthermore, the J-state interference is also significant in the polarization profile in the near wings of the H line (see right panel of Figure \ref{Fig_3_with_Hline_Unmag_HK_lines_diff_atoms_FALC}). Similar conclusions were found for the Mg~{\sc{ii}} h and k lines (see \citealt{Mg_II_HK_Belluzzi}, Figure 2).\\
 
    On the other hand, the amplitude of the polarization profiles of the IR triplet lines depends significantly on the treatment of the atomic system, as can be seen in Figure \ref{Fig_2_Unmag_IRT_lines_diff_atoms_FALC}. Such differences are due to how the flat-spectrum component of the emission vector is treated in the two cases \citep{Flat_spectrum}; for the multi-term atom, all transitions between the same two terms are assumed to be illuminated by the same radiation field, whereas a different radiation field can be considered for the transitions between different J levels for the multi-level atom. The radiation field treatment made when considering multi-level atoms can be problematic when the lines corresponding to a single term have a large separation in frequency, as in the case of the IR triplet. Unless otherwise noted, the results shown in the following sections of this work were obtained considering the 5L multi-level treatment. \\

\subsection{Impact of the metastable levels}\label{Subsection_Unmag_metastable}
\begin{figure}
        \centering
        \includegraphics[width=0.4\textwidth]{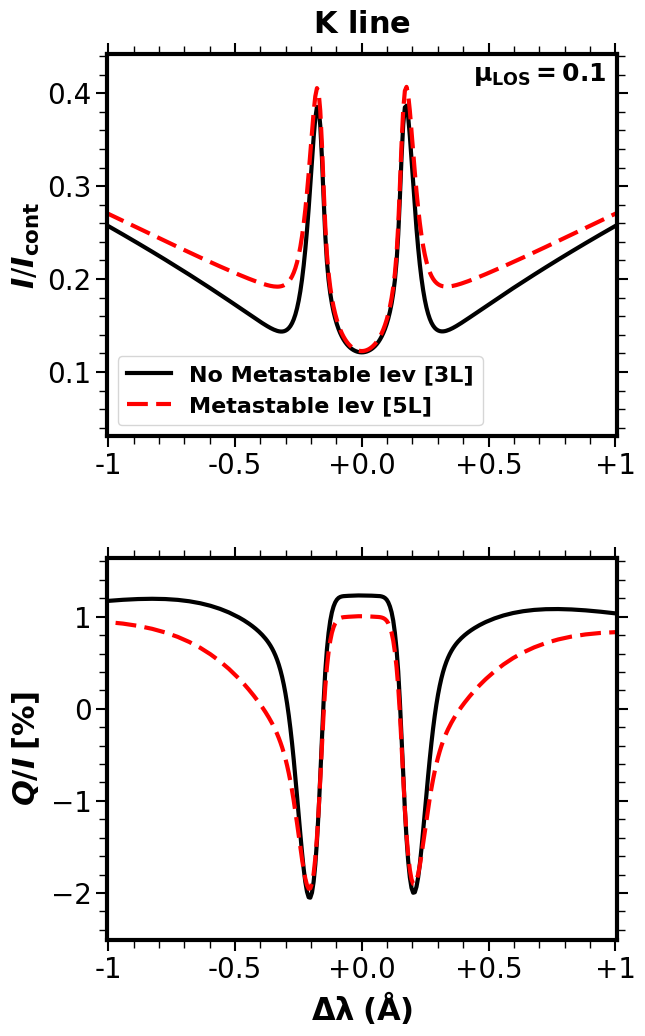}
        \caption{Stokes $I$, normalized to the continuum intensity $I_\mathrm{cont}$ (upper panel), and fractional linear polarization $Q/I$ profiles (lower panel) for the core and near-wing region of the K line as a function of wavelength distance to the center of the line. The considered redistribution treatment is PRD and the emergent profiles are shown for a near-the-limb line of sight, with $\mu=0.1$. The solid black and dashed red curves correspond to calculations for 3L (which ignores the metastable levels) and 5L (including the metastable levels) atomic models, respectively.}
        \label{Fig_4_Unmag_diff_atoms_K_muLOS_01_FALC}
    \end{figure}

    In this subsection, we investigate the impact of the $3d\, {}^2\mathrm{D}$ metastable levels on the K line, which are the lower levels of the Ca~{\sc{ii}} IR triplet. In Figure \ref{Fig_4_Unmag_diff_atoms_K_muLOS_01_FALC}, we compare the intensity and polarization profiles obtained considering the 5L (dashed red) and 3L (solid black) multi-level atomic models, where metastable levels are included as in the previous sections and excluded, respectively. The 3L case was computed with the upper and ground-level atomic populations fixed to those obtained from the solution of the 5L case, in order to specifically analyze the impact due to atomic polarization. We only show the K line because the IR triplet is not produced in the absence of metastable levels and the center of the H line cannot be linearly polarized due to scattering processes.\\

    The intensity profiles are in good agreement in the core region of the K line, but there is a discrepancy in the near wings. The inclusion of the metastable levels (5L atom) reduces the mean life-time of the upper level of the K line with respect to the 3L atom. This results in a generally smaller coherence fraction (the amount of coherent scattering events with respect to non-coherent emission) and, thereby, the intensity profiles in the line wings are enhanced. Furthermore, $Q/I$ is overestimated in the line-core region when metastable levels are not included. The atomic alignment of the $3d\, {}^2\mathrm{D}$ metastable term, connected to the $4p\,{}^2\mathrm{P}$ term via the IR triplet lines and collisional transitions, has an impact in the atomic polarization of the upper level of the K line, which is perceptible in the polarization profile of its core. The need to include the metastable levels in the atomic system to correctly model the polarization of the K line was already noted in Section 5 of \citet{TB_SPW_8}, although without accounting for PRD effects. There is also a clear discrepancy in the near wings of $Q/I$, although this is due to the already mentioned discrepancy in intensity, not as a result of direct impact on polarization. As shown in section \ref{Subsection_Unmag_atomic_models}, the profiles for the 5L and 3T models are almost the same in this frequency range, and we have checked that the same occurs for the 3L and 2T models, so the results are valid for both the multi-level and multi-term models. 

    \begin{figure*}[t]
        \centering
        \includegraphics[width=0.75\textwidth]{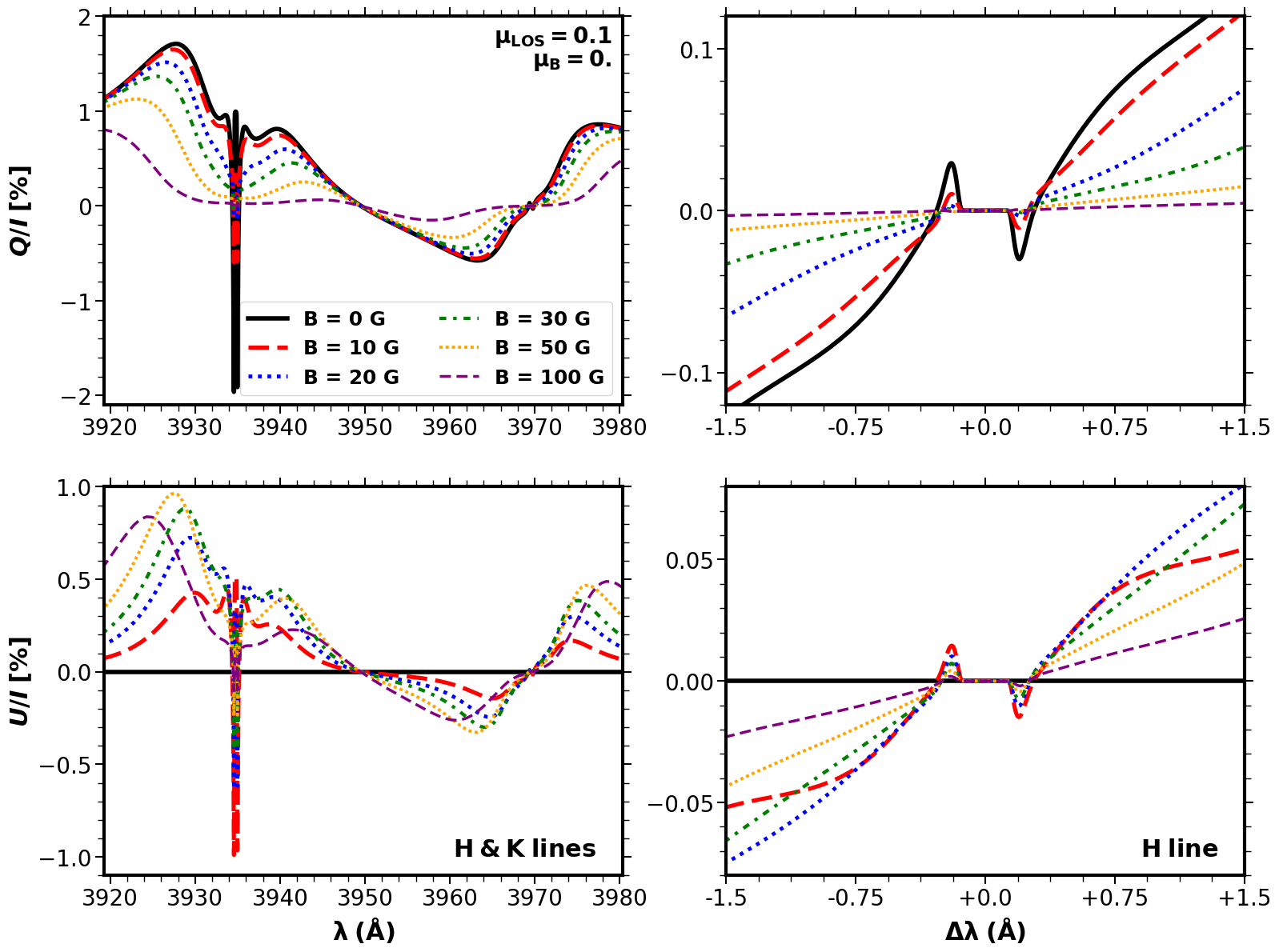}
        \caption{Fractional linear polarization $Q/I$ (upper panels) and $U/I$ (lower panels) profiles for the region around the H and K lines (left panels) and the core of the H line (right panels). The 3T multi-term atomic models are considered, so J-state interference is accounted for. The emergent profiles are shown for a LOS with $\mu=0.1$. The curves represent emergent profiles for syntheses considering a horizontal magnetic field ($\theta_{B} = 90^\circ$) with different strengths, indicated in the legend. The reference direction for positive Stokes $Q$ is the parallel to the nearest limb.}
        \label{Fig_3_Magnetized_HK_lines_diff_atoms_FALC}
    \end{figure*}

\section{The magnetized case} \label{Section_Mag}

In this section, we investigate the magnetic sensitivity of the lines under consideration. For this purpose, we set uniform magnetic fields with different strengths and orientations, and we analyze the Stokes profiles emerging at disk-center and near the limb.\\

In general, the qualitative results regarding the redistribution treatment and atomic model are the same as in the unmagnetized reference case. In the H and K lines, a PRD treatment is needed to correctly model the polarization profiles, and both the multi-level and multi-term atoms yield the same profiles in the line cores. However, the correct modeling of the profiles in the frequency range between the H and K lines requires a multi-term treatment to take J-state interference into account.\\

For the IR triplet lines, the polarization profiles overlap almost perfectly for all three redistribution cases, with slight differences for the CRD case. The deviations in the intensity wings of the IR triplet lines are still present when applying a full PRD treatment. Similarly to the unmagnetized case, the atomic model has a significant effect on the polarization profile amplitudes of the $8542 \:\text{\AA}$ and $8662 \: \text{\AA}$ lines.\\

\subsection{Effect of a horizontal magnetic field for a near-limb LOS ($\mu_{B} = 0$, $\mu = 0.1$)} \label{Subsection_Mag_mu01}
        
    In this subsection, we consider a uniform and horizontal magnetic field ($\theta_{B} = 90^\circ, \: \mu_{B} = 0$, parallel to the solar surface) with the azimuth selected to maximize its longitudinal component ($\chi_{B} = \chi = 0º$). We analyze the profiles of the emergent radiation for a LOS close to the limb, with $\mu = 0.1$.\\
    
    In order to study the magnetic sensitivity of the interference pattern in the UV doublet, the left panels in Figure \ref{Fig_3_Magnetized_HK_lines_diff_atoms_FALC} show the $Q/I$ (upper panel) and $U/I$ (lower panel) linear polarization profiles around the H and K lines for magnetic fields of different strength, ranging between $0$ and $100 \: \mathrm{G}$. In the right side panels, we show the same profiles around the core of the H line. The syntheses were computed considering a 3T atom (we accounted for J-state interference, which is necessary to correctly model the scattering polarization in this frequency range) and the IRCRD redistribution treatment. The emergent profiles are shown for a LOS with $\mu = 0.1$. Similarly to the h and k lines of Mg~{\sc{ii}} (see Figure 1 of \citealt{HanleRT}), the $Q/I$ profile in the range between the H and K lines, as well as in the near wings of H, is depolarized due to the magnetic field. For $B = 100 \: \mathrm{G}$, the $Q/I$ signal is close to zero in the frequency range between the H and K lines, although it increases to $\sim 1\%$ in the outer region. Magneto-optical effects produce a rotation and net depolarization of the emergent linear polarization, resulting in a significant $U/I$ signal when magnetic fields are present. Going further into the wings, the polarization profiles obtained in the presence of different magnetic fields show a better agreement, coinciding by $3880~\AA$ (on the blue side) and $4020~\AA$ (on the red side), due to the fact that the magnetic sensitivity diminishes as the line becomes less opaque. \\

    \afterpage{
    \clearpage
    \begin{figure*}[h]
        \centering
        \includegraphics[width=0.75\textwidth]{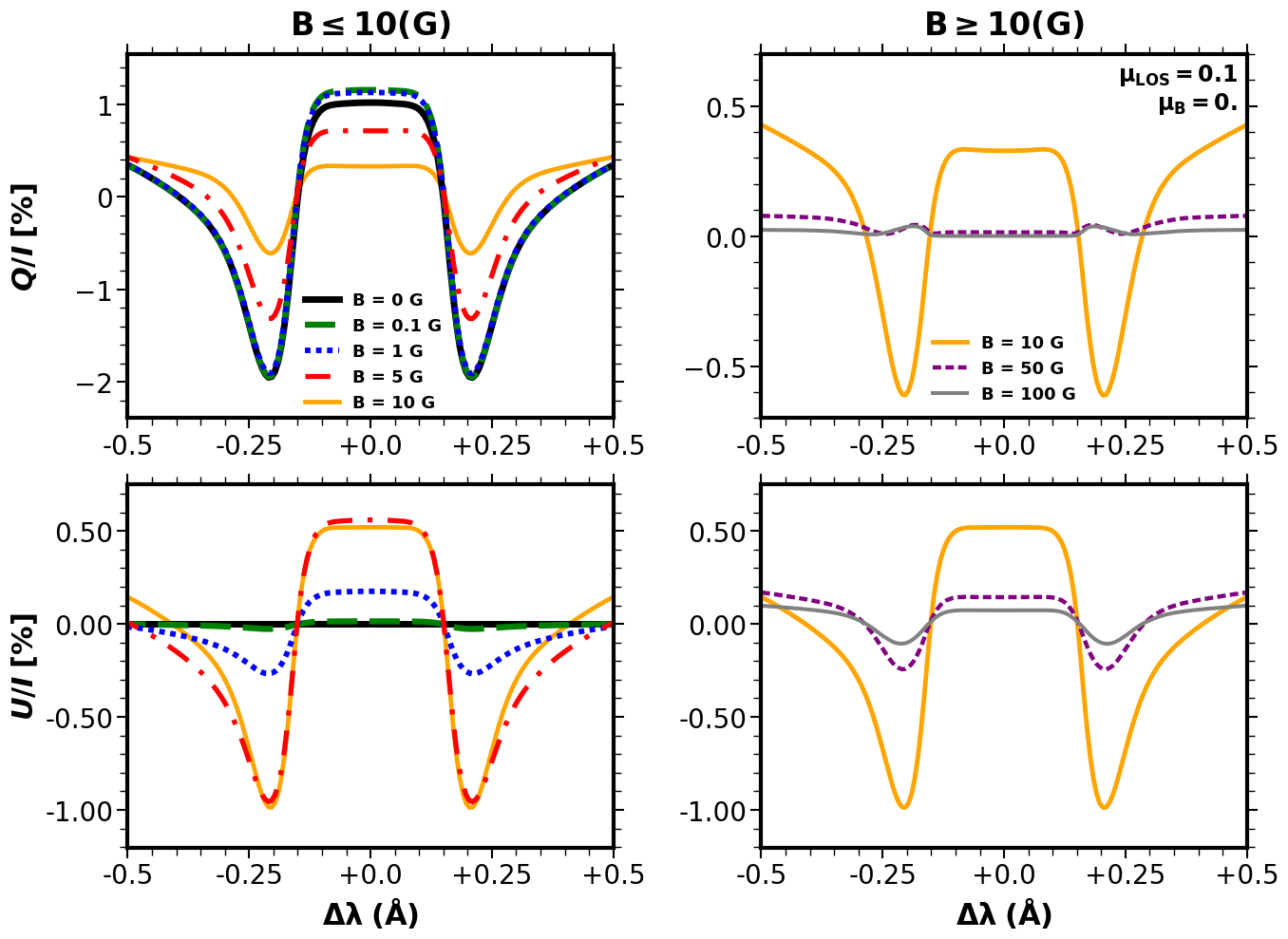}
        \captionof{figure}{Fractional linear polarization profiles, $Q/I$ (upper panels) and $U/I$ (lower panels), for the core and near-wing region of the K line, for a line of sight with $\mu=0.1$. The curves represent emergent profiles for syntheses considering a uniform and horizontal magnetic field with different strengths, indicated in the legend. Results for $B = 0 - 10 \: \mathrm{G}$ are shown in the left panels, whereas results for $B = 10 - 100 \: \mathrm{G}$ are shown in the right ones. The $B=10 \: \mathrm{G}$ case is represented in all the panels as a solid orange curve in order to allow for comparison. Notice that the ordinate scale is not the same for all panels. The reference direction for positive Stokes $Q$ is the parallel to the nearest limb.}
        \label{Fig_6_Mag_K_line_diff_B_FALC}
    \end{figure*}
        
    \begin{figure*}[b]
        \centering
        \includegraphics[width=1.\textwidth]{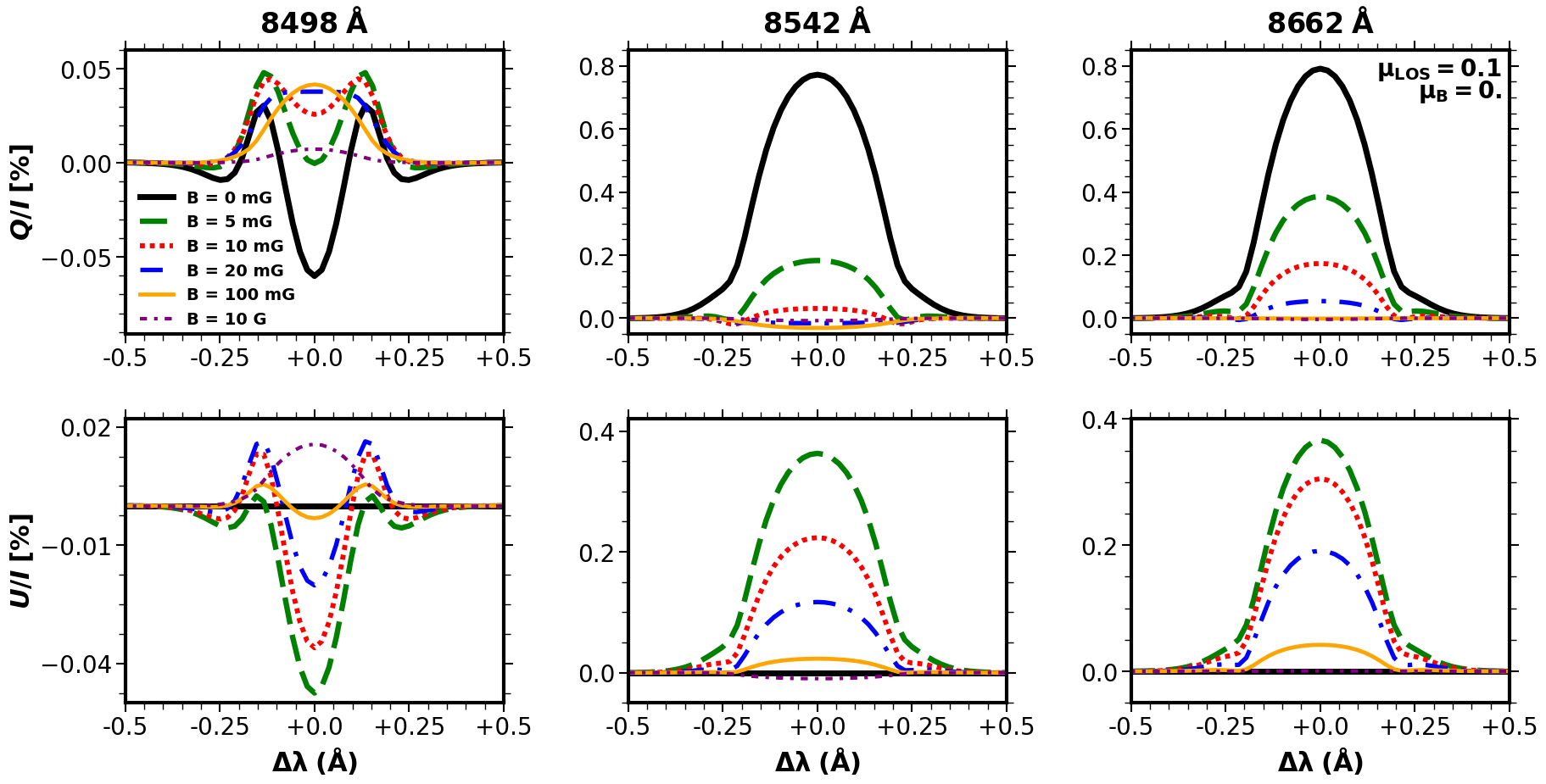}
        \captionof{figure}{Fractional linear polarization profiles, $Q/I$ (upper panels) and $U/I$ (lower panels), for the core and near-wing region of the IR triplet lines for a LOS with $\mu=0.1$; from left to right, the $8498 \: \AA$, $8542 \: \AA$, and $8662 \: \AA$ lines. The curves represent emergent profiles for syntheses with different uniform and horizontal magnetic fields with strengths in the $B = 0 -10 \: \mathrm{G}$ range, as indicated in the legend. The reference direction for positive Stokes $Q$ is the parallel to the nearest limb.}
        \label{Fig_5_Mag_IRT_diff_B_FALC}
    \end{figure*}
    \clearpage
    }
    
    Figures \ref{Fig_6_Mag_K_line_diff_B_FALC} and \ref{Fig_5_Mag_IRT_diff_B_FALC} show the $Q/I$ and $U/I$ linear polarization profiles in the core and near-wing regions, for the K line and the IR triplet lines, respectively. These profiles were computed with an IRCRD treatment and a 5L atomic model for a LOS with $\mu = 0.1$, and the curves represent results for different strengths of the magnetic field, ranging from values of milligauss ($B = 5 \: \mathrm{mG}$) up to a hundred gauss ($B=100 \: \mathrm{G}$), as well as the unmagnetized reference case. \\

    The left panels of Figure \ref{Fig_6_Mag_K_line_diff_B_FALC} display the profiles for the K line corresponding to syntheses in the $0-10 \: \mathrm{G}$ range, whereas the right panels show results for larger field strengths ($B \ge 10 \: \mathrm{G}$). For field strengths between $0.1$ and $1 \: \mathrm{G}$, we find an increase of $Q/I$ in the core of the K line with respect to the zero-magnetic-field profiles (see left panels of Figure \ref{Fig_6_Mag_K_line_diff_B_FALC}). Even though, for these field strengths, the Hanle effect has no significant impact on the upper level of the transition ($4p^{2} \mathrm{P}^{\mathrm{o}}_{3/2}$), it does operate significantly on the $3d\, {}^2\mathrm{D}$ metastable levels, modifying their atomic polarization, which has its feedback on the atomic polarization of the $4p^{2} \mathrm{P}^{\mathrm{o}}_{3/2}$ upper level (see \citealp{TB_SPW_8}). We verified that this effect disappears when the metastable levels are not included in the atomic model. For stronger magnetic fields, the Hanle effect also operates on the upper level of the K line. The amplitude of the $Q/I$ profile decreases with increasing field strength until it is close to saturation for field strengths beyond $50 \: \mathrm{G}$ (see right panels of Figure \ref{Fig_6_Mag_K_line_diff_B_FALC}). The $U/I$ profile is zero in the unmagnetized case, but its amplitude increases with the field strength until reaching a maximum at around $5-10 \: \mathrm{G}$, mainly due to Hanle rotation. For stronger magnetic fields the amplitude decreases due to Hanle depolarization. \\

    The profiles for the IR triplet corresponding to magnetic fields strengths ranging from milligauss to 10 G are shown in Figure \ref{Fig_5_Mag_IRT_diff_B_FALC}. As noted above, the lower levels of these transitions are the metastable levels, whose atomic polarization is sensitive to the Hanle effect for sub-gauss strengths. Selective absorption of polarization components plays a very important role in the scattering polarization of the IR triplet lines; indeed, because the upper level $4p^{2} \mathrm{P}^{\mathrm{o}}_{1/2}$ cannot have atomic alignment, the scattering polarization of the $8662 \: \AA$ line is entirely due zero-field dichroism \citep{Dichroism, MSTB_2010}. As a result, sub-gauss magnetic fields have a significant impact on the linear polarization profiles of the IR triplet lines. The $8542 \: \AA$ and $8662 \: \AA$ lines show very similar qualitative behaviors: $Q/I$ rapidly decreases with respect to the unmagnetized case when a field is present, almost reaching saturation for $B=100 \: \mathrm{mG}$. The $U/I$ profiles increase to measurable amplitudes ($\sim 0.4 \: \%$) for very small field strengths ($B = 5 \: \mathrm{mG}$) and decrease steadily when increasing the strength beyond this value. As evidenced by the difference between the curves corresponding to $B=100 \: \mathrm{mG}$ and $B=10 \: \mathrm{G}$ in the left panel, magnetic fields slightly stronger than $10 \: \mathrm{G}$ are required in order to reach saturation in the $8498 \: \AA$ line. \\

    \subsection{Effect of a horizontal magnetic field in forward \protect\\ scattering ($\mu_{B} = 0$, $\mu = 1$)}\label{Subsection_Mag_mu1}
   
    We now analyze the same magnetic geometrical setup as in section \ref{Subsection_Mag_mu01} (horizontal and uniform field), but considering the profiles for the radiation emerging at disk center, i.e., with $\mu = 1$. For this line of sight, considering 1D static atmospheric models, linear scattering polarization can only be produced when a magnetic field breaks the axial symmetry. We take the reference direction for positive $Q/I$ to be perpendicular to the direction of the magnetic field. Clearly, $U/I$ is zero for the considered geometry.\\

    \begin{figure}[H]
        \centering
        \includegraphics[width=0.45\textwidth]{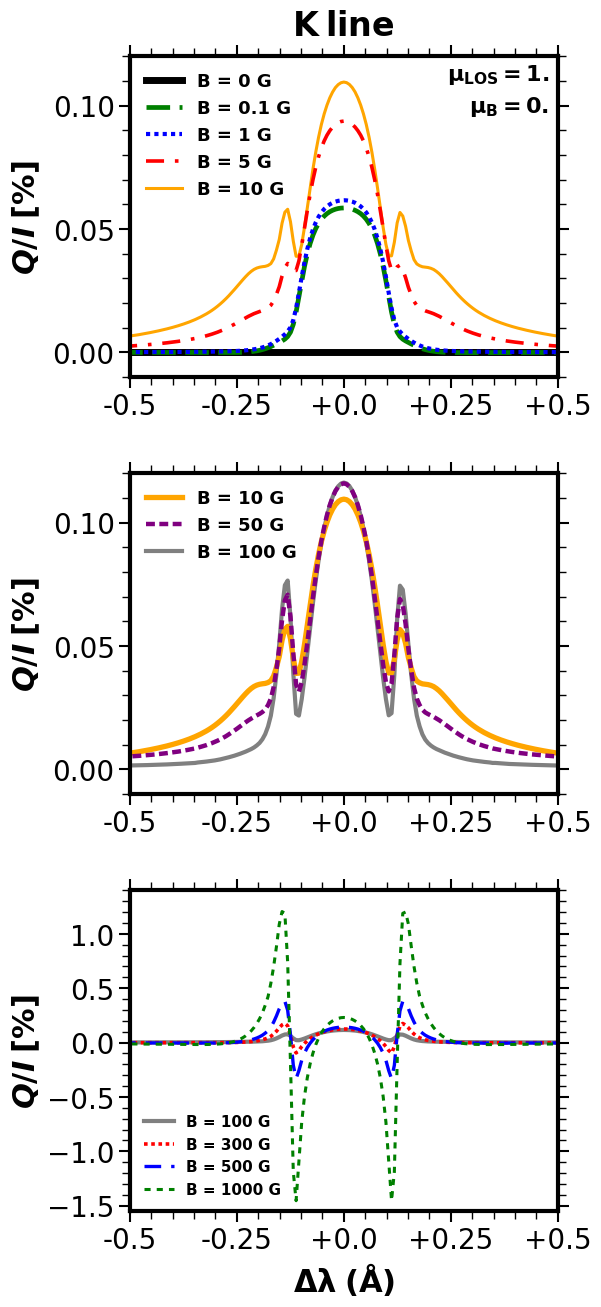}
        \captionof{figure}{Fractional linear polarization profiles $Q/I$ for the core and near-wing region of the K line for the disk-center LOS, with $\mu=1$. The curves represent emergent profiles for syntheses with different magnetic field strengths, as indicated in the legend. Results for $B = 0 - 10 \: \mathrm{G}$ are shown in the upper panel, $B = 10 - 100 \: \mathrm{G}$ in the central panel, and $B = 100 - 1000 \: \mathrm{G}$ in the lower one. The $B=10 \: \mathrm{G}$ case is represented in the upper and central panel as a solid orange curve in order to facilitate the comparison. Likewise, the $B=100 \: \mathrm{G}$ is shown in the central and lower panels as a solid grey curve. The reference direction for positive Stokes $Q$ is the perpendicular to the magnetic field.}
        \label{Fig_10_Mag_K_line_diff_B_FALC}
    \end{figure}

    \begin{figure*}
        \centering
        \includegraphics[width=1\textwidth]{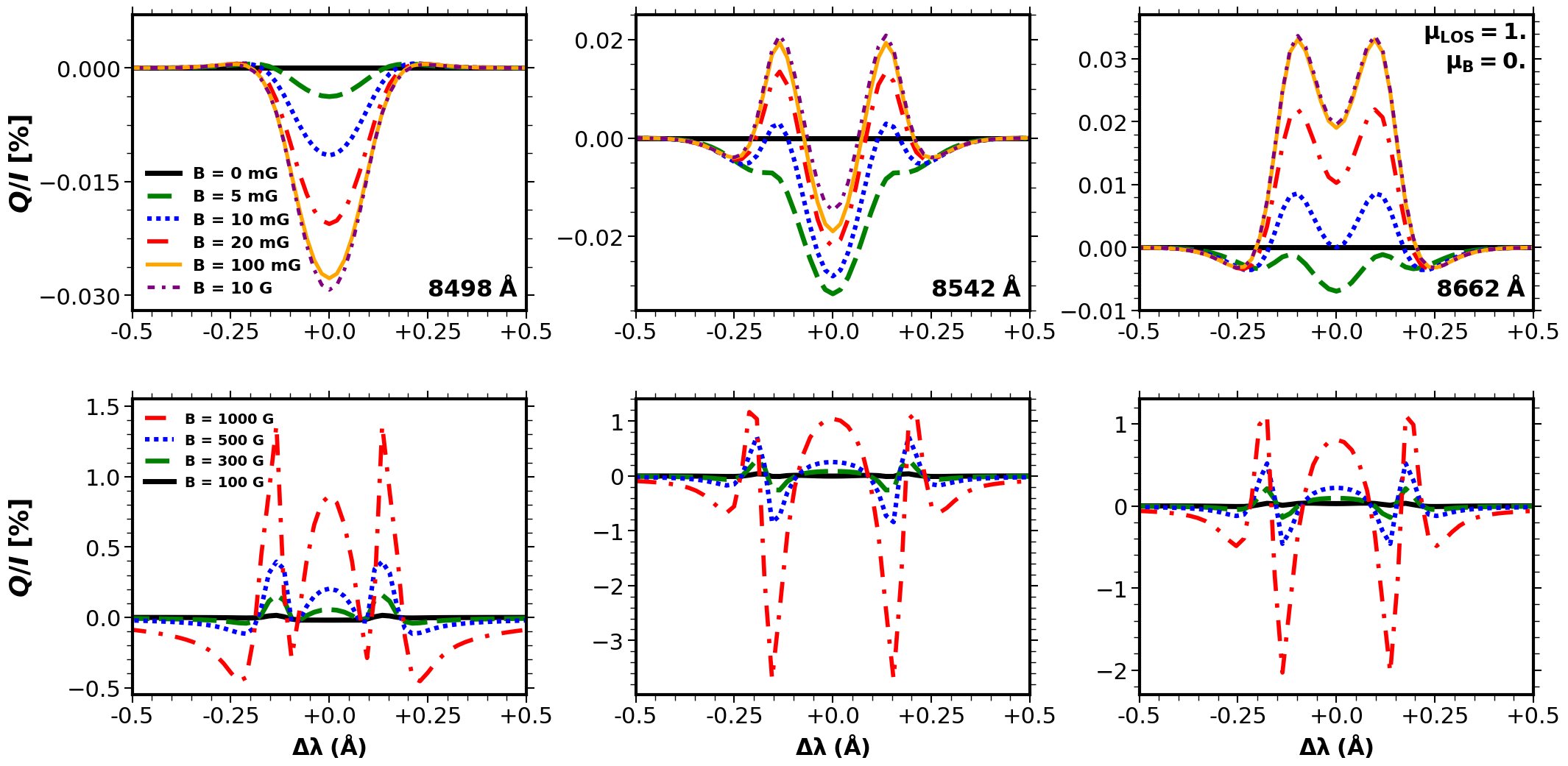}
        \captionof{figure}{Fractional linear polarization profile $Q/I$ for the core and near-wing region of the IR triplet lines for the disk-center LOS ($\mu = 1$); from left to right, the $8498 \: \AA$, $8542 \: \AA$, and $8662 \: \AA$ lines. The curves in the upper panels represent emergent profiles for syntheses with uniform and horizontal magnetic fields with strengths in the $B = 0 -10 \: \mathrm{G}$ range, as indicated in the legend, whereas lower panels show results in the $B = 100 -1000 \: \mathrm{G}$ range. The reference direction for positive Stokes $Q$ is the perpendicular to the magnetic field.}
        \label{Fig_9_Mag_IRT_diff_B_FALC}
    \end{figure*}
    
    \begin{figure}
        \centering
        \includegraphics[width=0.4\textwidth]{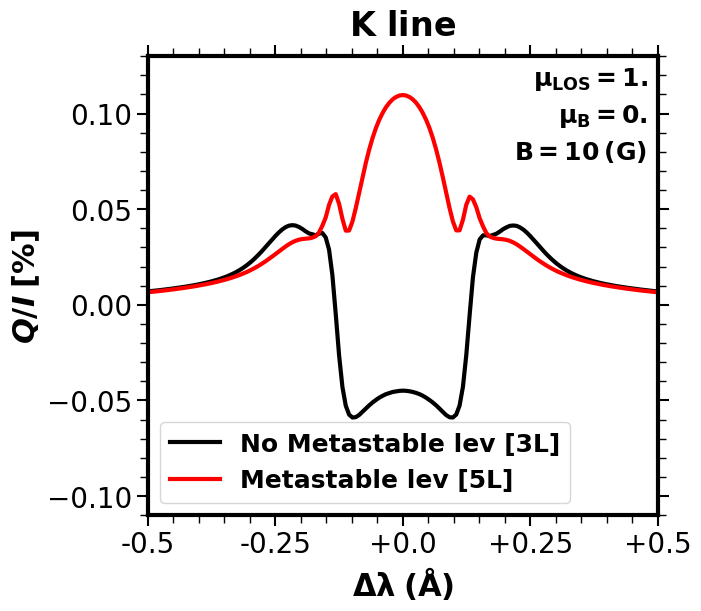}
        \caption{Fractional linear polarization $Q/I$ profile for the core and near-wing region of the K line as a function of wavelength distance to the center of the line. The considered redistribution treatment is PRD, a horizontal magnetic field of $B=10 \: \mathrm{G}$ is present and the emergent profiles are shown for the center of the solar disk, with $\mu=1$. The solid black and red curves correspond to calculations for 3L (neglecting metastable levels) and 5L (including the metastable levels) atomic models, respectively. The reference direction for positive Stokes Q is the perpendicular to the horizontal magnetic field.}
        \label{Fig_14_Mag_diff_atoms_K_muLOS_1_FALC}
    \end{figure}
    
    Figure \ref{Fig_10_Mag_K_line_diff_B_FALC} shows the $Q/I$ profiles for the K line, computed for uniform and horizontal magnetic fields, with the upper and middle panels corresponding to the same field strengths considered in Figure \ref{Fig_6_Mag_K_line_diff_B_FALC}. Despite not directly affecting the upper levels of the K line, sub-gauss magnetic fields already give rise to significant polarization signals by modifying the atomic polarization of the mestastable levels, as pointed out in \ref{Subsection_Mag_mu01}. When accounting for partial frequency redistribution, symmetric peaks appear on either side of the K line, at $\Delta \lambda \sim 0.15 \: \AA$ from line center, for fields stronger than $B = 5 \: \mathrm{G}$. The onset of Hanle saturation begins to be appreciable at $B = 10 \: \mathrm{G}$, and it is completely reached at $B = 50 \: \mathrm{G}$. On the other hand, the amplitude of the side peaks increases further for even stronger fields. We attribute this to the Zeeman effect, whose contribution to the linear polarization amplitude increases with the square of the transverse component of the magnetic field. We provide support for this hypothesis in Appendix \ref{Appendix_forward_scatt_trans_Zeeman}. These peaks are not produced when the CRD treatment is made, in which case the shape of the $Q/I$ profile for all considered field strengths is similar to that found for $B = 0.1$ and $1 \: \mathrm{G}$. Moreover, as shown in the lower panel of Figure \ref{Fig_10_Mag_K_line_diff_B_FALC}, the linear polarization signals are essentially dominated by the Zeeman effect in the presence of magnetic fields on the order of hundreds of gauss. The contribution from the Zeeman effect scales with the square of the transverse component of the magnetic field and the $Q/I$ amplitude reaches $\sim 1.5\%$ for $B=1000 \: \mathrm{G}$.\\
    
    The polarization amplitude produced by the Hanle effect in the IR triplet increases with field strength until $B\sim100 \: \mathrm{mG}$, as can be seen in the upper panels of Figure \ref{Fig_9_Mag_IRT_diff_B_FALC}. Around this strength, the metastable levels (i.e., the lower levels of the transitions) are near Hanle saturation; further increases in field strength have no appreciable impact on the resulting profiles, as evidenced by the profiles obtained for $B = 100 \: \mathrm{mG}$ and $B = 10 \: \mathrm{G}$, which are almost indistinguishable. We recall that the magnetic field modifies the atomic polarization in the metastable levels and, thus, the dichroism for the lines of the IR triplet, which in turn impacts their linear polarization profiles (see \citealt{MSTB_2010}). At saturation, we find that the fractional linear polarization of the IR triplet lines at $\mu = 1$ is still below $0.05\%$. For magnetic fields stronger than about a hundred gauss, the linear polarization is again dominated by the Zeeman effect. For $B = 1000 \: \mathrm{G}$, the amplitude of the characteristic Zeeman $Q/I$ profiles of the IR triplet reach $\sim 3\%$.\\

    In the magnetic case, including the metastable levels in the atomic model is also necessary to correctly model the polarization profiles of the K line, as shown in Figure \ref{Fig_14_Mag_diff_atoms_K_muLOS_1_FALC}. Here, similarly to Figure \ref{Fig_4_Unmag_diff_atoms_K_muLOS_01_FALC}, we show the resulting $Q/I$ polarization profile for 3L and 5L multi-level atomic models (excluding and including metastable levels, respectively), but in forward scattering and for the $B=10 \: \mathrm{G}$ case. These results are analogous to the strongest field case in Figure 8 of \cite{TB_SPW_8}, but taking PRD effects into account\footnote{Note that in Figure 8 of \cite{TB_SPW_8} the positive reference direction for Stokes Q is the parallel to the horizontal magnetic field.}.\\

\vspace{-0.3cm}
\subsection{Effect of a vertical magnetic field in near-limb case ($\mu_{B} = 1$, $\mu = 0.1$)} \label{Subsection_Mag_vert_mu01}

    \begin{figure}[!t]
        \centering
        \includegraphics[width=0.35\textwidth]{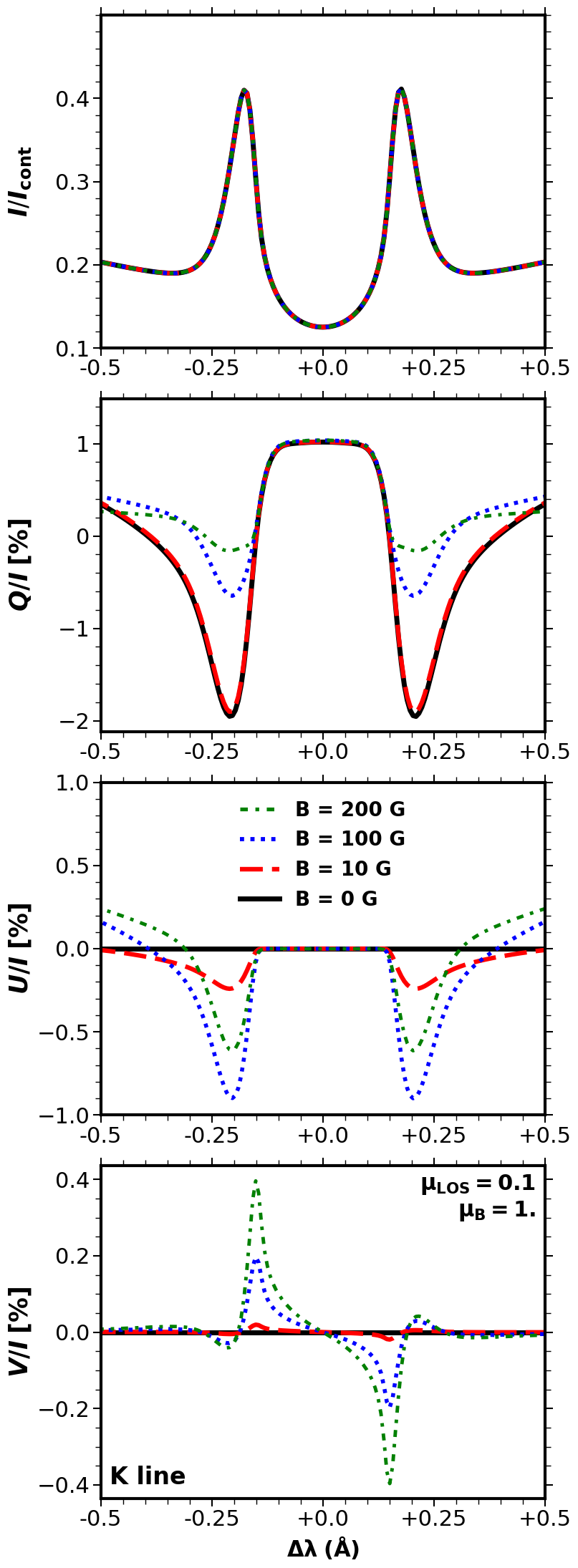}
        \caption{Stokes profiles for the core and near-wing region of the K line for a LOS with $\mu=0.1$: $I/I_\mathrm{cont}$ (first panel), $Q/I$ (second panel), $U/I$ (third panel), and $V/I$ (fourth panel). The curves represent the emergent profiles considering uniform and vertical magnetic fields with different strengths, indicated in the legend. The reference direction for positive Stokes Q is the parallel to the nearest limb.}
        \label{Fig_17_Vertical_B_mu_01_K_line}
    \end{figure}

    \begin{figure}[!t]
        \centering
        \includegraphics[width=0.35\textwidth]{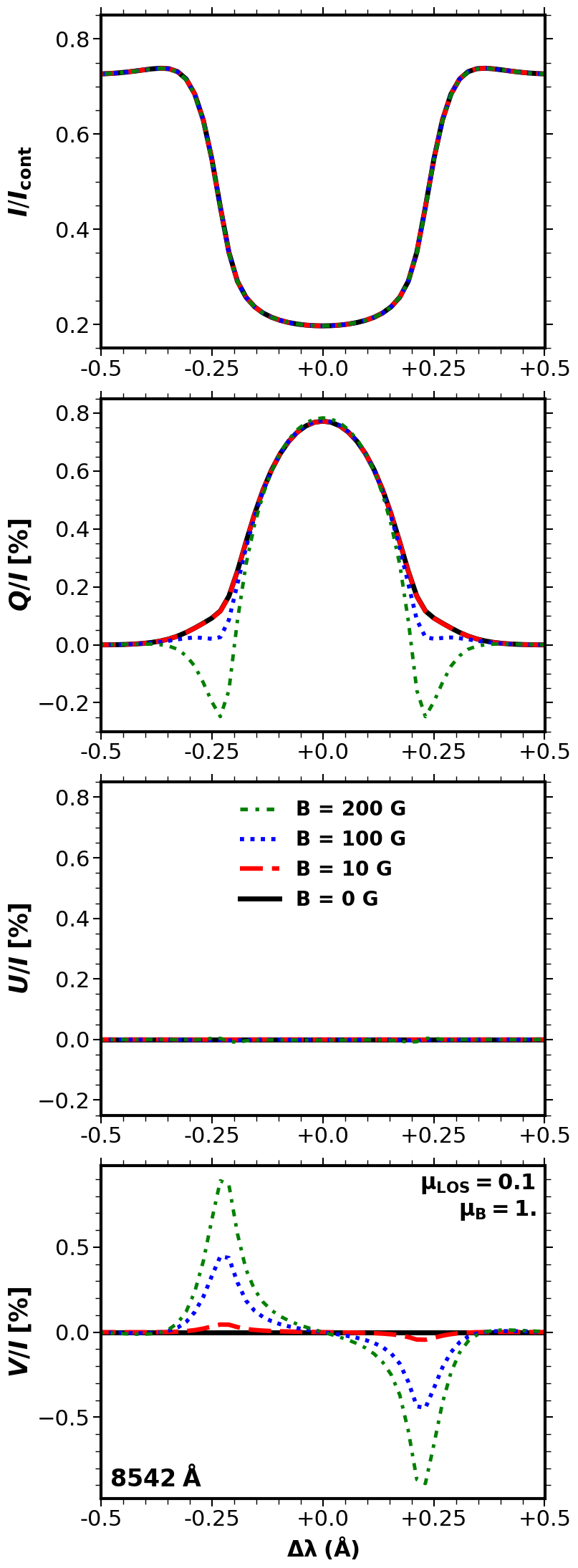}
        \caption{Stokes profiles for the core and near-wing region of the $8542 \: \AA$ line for a LOS with $\mu=0.1$: $I/I_\mathrm{cont}$ (first panel), $Q/I$ (second panel), $U/I$ (third panel), and $V/I$ (fourth panel). The curves represent the emergent profiles considering uniform and vertical magnetic fields with different strengths, indicated in the legend. The reference direction for positive Stokes Q is the parallel to the nearest limb.}
        \label{Fig_17_Vertical_B_mu_01_8542_A}
    \end{figure}
    
    In the present subsection, we impose a uniform vertical magnetic field ($\theta_{B} = 0^\circ, \: \mu_{B} = 1$) and we analyze results for a near-limb LOS ($\mu = 0.1$). The emergent profiles, considering various field strengths up to $200 \: \mathrm{G}$, are shown in Figures \ref{Fig_17_Vertical_B_mu_01_K_line} and \ref{Fig_17_Vertical_B_mu_01_8542_A} for the K and the $8542 \: \AA$ lines, respectively. Appreciable circular polarization signals are produced in both lines despite the small LOS component of the magnetic field, whereas the intensity profiles are unaffected even by the strongest fields considered here. \\
    
    For the considered geometry, the magnetic field is parallel to the symmetry axis of the radiation field and, thus, the Hanle effect does not operate on the atomic level polarization. Thus, the polarized radiation in the line core regions is unaffected by the magnetic field (under the AA approximation), as can be appreciated in the second and third panels. However, the polarization in the near-wing region of the K line (see Figure \ref{Fig_17_Vertical_B_mu_01_K_line}) is clearly modified by magneto-optical effects that rotate the polarization signals, reducing the amplitude in $Q/I$ while increasing it in $U/I$ (\citealt{Mg_II_Ernest, HanleRT}). This gives rise to $U/I$ signals with amplitudes up to $-1 \%$ around $\Delta \lambda \sim 0.2$ from the line center. The total linear polarization fraction is reduced by the magneto-optical effects (see appendix A in \citealt{Alsina_2018}), which can be appreciated comparing the $100 \: \mathrm{G}$ and $200 \: \mathrm{G}$ cases. In the AD case, the magnetic field can have an impact in the core region, as discussed in section 3.1. of \citet{Tanausu_AD} for the Mg~{\sc{ii}} k line. As stated before, this will be addressed in a future work. \\

    On the other hand, the $8542 \: \AA$ line is unaffected by the magneto-optical effects due to the lack of substantial scattering polarization signals in its wings. As shown in the third panel of Figure \ref{Fig_17_Vertical_B_mu_01_8542_A}, the amplitude of the $U/I$ profile is zero because there are no effective mechanisms to rotate the polarization signals. In the presence of a vertical magnetic field, the linear polarization profiles in the $8542 \: \AA$ line (and the rest of the IR triplet lines) are solely modified by the Zeeman effect due to the transverse component of the field. As a result, we only expect noticeable changes in the linear polarization profiles compared to the unmagnetized case to occur for field strengths of hundreds of gauss, as shown by the green dotted curve in the second panel, corresponding to $B = 200 \: \mathrm{G}$.\\

\section{The circular polarization profiles} \label{Section_circular_pol}

\begin{figure}
        \centering
        \includegraphics[width=0.4\textwidth]{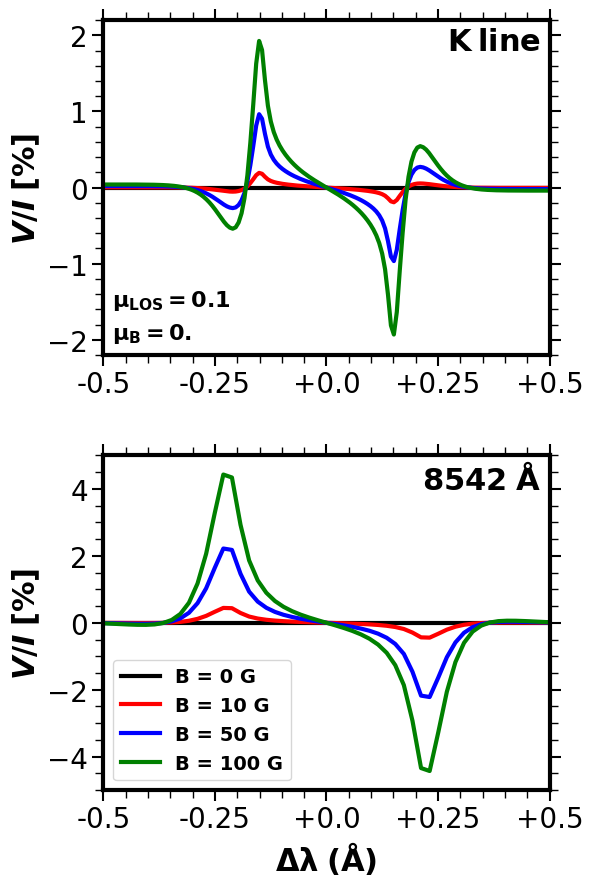}
        \caption{Fractional circular polarization profiles $V/I$ for the core and near-wing region of the K (upper panel) and $8542 \: \AA$ (lower panel) lines, for a close to the limb LOS with $\mu=0.1$. The curves represent emergent profiles for syntheses with uniform horizontal magnetic fields with different strengths: $B = 0 \: \mathrm{G}$ (black), $10 \: \mathrm{G}$ (red), $50 \: \mathrm{G}$ (blue), and $100 \: \mathrm{G}$ (green).}
        \label{Fig_7_8_Unmag_K_8542_lines_diff_B_Stokes_V_FALC}
    \end{figure}

    In this section, we show the circular polarization profiles obtained through the calculations discussed in section \ref{Section_Mag}. Figure \ref{Fig_7_8_Unmag_K_8542_lines_diff_B_Stokes_V_FALC} shows the $V/I$ circular polarization profiles for a near-limb LOS ($\mu=0.1$), resulting from IRCRD calculations considering a 5L model and uniform horizontal magnetic fields of various strengths. We find that the amplitude of the circular polarization increases linearly within the considered range (up to $300 \: \mathrm{G}$), while the shape of the profiles remains the same.\\
    
    The figure shows the K and $8542 \: \AA$ lines, whose circular polarization profiles we consider to be representative of those of the various lines of the UV doublet and IR triplet, respectively. Indeed, we verified the qualitative similarity between the profiles for the lines belonging to the same multiplet. We also verified that the results coincide when considering a 3T multi-term atom instead of a 5L multi-level one, or a PRD instead of an IRCRD treatment. The profiles for the vertical magnetic field case are also analogous when observing at the disk center.\\

\begin{figure}
        \centering
        \includegraphics[width=0.4\textwidth]{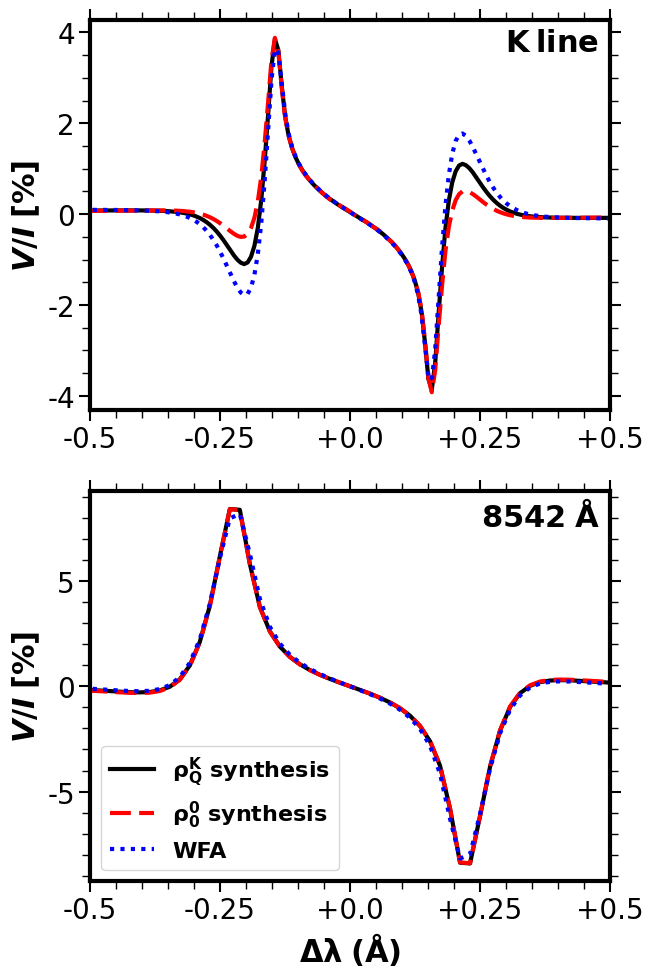}
        \caption{Fractional circular polarization profiles $V/I$ for the core and near-wing region of the K (upper panel) and $8542 \: \AA$ (lower panel) lines. The emergent profiles are shown for a line of sight with $\mu=0.1$. The imposed $200 \: \mathrm{G}$ magnetic field is horizontal, height-independent, and quasi-parallel ($\mu_{B}=0, \: \mu = 0.1$) to the observer. The curves represent the synthetic profiles with (solid black) and without (dashed red) atomic level polarization, and the WFA result (dotted blue).}
        \label{Fig_11_Mag_K_8542_V_pol_nopol}
    \end{figure}

    Figure \ref{Fig_11_Mag_K_8542_V_pol_nopol} displays the synthetic $V/I$ profiles obtained in the $B = 200 \: \mathrm{G}$ case for a LOS with $\mu = 0.1$, alongside the profiles obtained neglecting atomic level polarization and the anisotropy and symmetry breaking of the pumping radiation field. This was achieved in HanleRT-TIC by setting the atomic ($\rho^K_Q$) and radiation field ($J^K_Q$) multipoles with $K>0$ to zero at each iteration. Thus, the resulting polarization patterns are exclusively due to the Zeeman effect. We discuss the linear polarization in Appendix \ref{Appendix_forward_scatt_trans_Zeeman}. As for the circular polarization patterns, the results for the $8542 \: \AA$ line obtained both accounting for atomic level polarization and neglecting it are in excellent agreement (we verified that this is also the case for the other lines of the IR triplet). For the K (and the H) line, the two syntheses match in the core region, but the amplitude of the outer lobes is underestimated when the atomic polarization is not accounted for. This is similar to the findings for the Mg~{\sc{ii}} h \& k lines in \cite{HanleRT_TIC} (see also \citealt{HanleRT, Mg_II_Ernest}).\\

    A circular polarization profile obtained from the application of the weak field approximation (WFA) to the intensity profile is also shown in Figure \ref{Fig_11_Mag_K_8542_V_pol_nopol}. The $8542 \: \AA$ line (lower panel) shows an excellent agreement between all the curves. Although the three curves also overlap in the core region of the K line (upper panel), there are discrepancies in their outer lobes, where the application of the WFA overestimates the circular polarization amplitude of the synthesis. This may lead to an underestimation in the longitudinal field component inferred through the WFA \citep{Mg_II_Ernest, HanleRT}. Indeed, in order to infer the LOS component of the magnetic field, we applied a least-squares fit to the $I$ and $V$ profiles of the K line for the general case in which atomic polarization was accounted for \citep[see also][]{Centeno_WFA}. This yields a field strength of $B_\mathrm{{LOS}}\mathrm{(WFA)} = 180.2 \: \mathrm{G}$, whereas the LOS projection of the magnetic field imposed in the synthesis was $B_\mathrm{{LOS}}\mathrm{(True)} = 199 \: \mathrm{G}$.\\

\section{Summary and concluding remarks} \label{Section_Conclusions}
The Ca~{\sc{ii}} spectral lines contain valuable thermodynamic and magnetic information of the solar chromosphere. In the chosen semi-empirical model atmosphere, the line cores of the IR triplet form at about $\sim 1000$~km, whereas the UV doublet does so around $\sim 1500$~km (see Figure 3 of \citealt{MS_2014}), thus encoding information from an extended height range comprising the middle and upper chromosphere. We carried out numerical syntheses using the spectral synthesis module of the HanleRT-TIC radiative transfer code in order to investigate the physical mechanisms required for an appropriate modeling of Ca~{\sc{ii}}, as well as the magnetic sensitivity of the resonance and subordinate spectral lines.\\

Whereas the PRD treatment is necessary to model the intensity and polarization signals in the near wings of the H and K lines, we find that the lines of the IR triplet can be suitably treated under the CRD  approximation, which introduces only a minor error. This conclusion validates the CRD multi-level investigations of \cite{Dichroism, MSTB_2010}, who studied the polarization of the Ca~{\sc{ii}} IR triplet and its magnetic sensitivity, accounting for atomic level polarization in all levels. The polarization profiles of the five lines can be suitably modeled simultaneously considering the IRCRD treatment discussed in the main text. However, the full PRD treatment introduces noticeable differences in the wings of the intensity profiles (see Figure \ref{Fig_1_Unmag_all_lines_diff_RD_FALC}).\\

Regarding the treatment of the atomic system, the scattering polarization in the core region of the H and K lines can be correctly modeled with either a multi-level or a multi-term atom. However, the polarization pattern in the range between the two lines is severely affected by J-state interference, which can only be taken into account in a multi-term model (see Figure \ref{Fig_3_with_Hline_Unmag_HK_lines_diff_atoms_FALC}). \citet{Mg_II_HK_Belluzzi} found analogous results for the resonant doublet of Mg~{\sc{ii}}. In either case, it is necessary to include the metastable levels in the atomic model to correctly treat the line-core linear polarization of the H and K lines, because the atomic polarization in such levels has an impact on that of the upper levels (see \citealt{TB_SPW_8}, and Figures \ref{Fig_4_Unmag_diff_atoms_K_muLOS_01_FALC} and \ref{Fig_14_Mag_diff_atoms_K_muLOS_1_FALC}). Regarding the IR triplet, J-state interference does not have a significant impact on their linear polarization patterns. It is important to note that considering the multi-term model in calculations of the Stokes profiles of the IR triplet leads to an overestimation of the linear polarization of these lines, because the flat-spectrum component of the emission vector is computed assuming the same radiation field for the three lines, despite their spectral separation (see Figure \ref{Fig_2_Unmag_IRT_lines_diff_atoms_FALC}). We find that these results hold both in the unmagnetized case and for the magnetic fields considered in this work.\\

We also investigated the sensitivity of the linear polarization of the Ca~{\sc{ii}} lines to uniform horizontal and vertical magnetic fields. The atomic polarization of the metastable levels is sensitive to magnetic fields in the $5-100 \: \mathrm{mG}$ range, which, when present, substantially reduce the amplitude of the IR triplet lines near the limb (see \citealt{MSTB_2010}, and Figure \ref{Fig_5_Mag_IRT_diff_B_FALC}). Such sub-gauss magnetic fields also lead to an increase of the linear polarization in the core of the K line as a consequence of the transfer of atomic polarization from the metastable to the upper levels via radiative and collisional transitions. Fields stronger than $5 \: \mathrm{G}$ directly modify the atomic polarization of the upper levels through the Hanle effect, leading to a depolarization and rotation of the linear polarization in the core of the K line, which intensifies with field strength until reaching saturation at $\sim 50 \: \mathrm{G}$ (see Figure \ref{Fig_6_Mag_K_line_diff_B_FALC}). In addition, the polarization signal in the spectral range between the H and K lines, which is dominated by PRD and J-state interference, is rotated and reduced in the presence of magnetic fields stronger than $10 \: \mathrm{G}$ due to magneto-optical effects (see Figure \ref{Fig_3_Magnetized_HK_lines_diff_atoms_FALC}).\\

For the disk-center LOS ($\mu = 1$), non-zero linear polarization signals arise in the presence of an inclined magnetic field, which is the only way to break axial symmetry in plane-parallel static model  atmospheres. When the magnetic field is horizontal, the forward-scattering linear polarization signals of the IR triplet appear for sub-gauss fields and increase with the field strength until around $100 \: \mathrm{mG}$, for which Hanle saturation is reached for the metastable levels (see Figure \ref{Fig_9_Mag_IRT_diff_B_FALC}). For the K line, sub-gauss magnetic fields already produce a polarization signal due to the aforementioned transfer of atomic polarization between the metastable and the upper levels. Beyond $5 \: \mathrm{G}$, scattering polarization peaks appear on either side of the K line if PRD effects are taken into account for the UV doublet, due to the Hanle effect that operates on its upper level. For larger field strengths, the linear polarization amplitude of the peaks continues increasing, even after Hanle saturation is reached for the upper level (see Figure \ref{Fig_10_Mag_K_line_diff_B_FALC}). This is a consequence of the Zeeman effect, which produces a significant contribution when the transverse component of the magnetic field is large, as discussed in Appendix ~\ref{Appendix_forward_scatt_trans_Zeeman}. For this geometry, the Zeeman effect begins to dominate the linear polarization signal of the K line at field strengths of several hundred gauss, whereas for the IR triplet this occurs already at a field strength of about one hundred gauss. \\

In the presence of a vertical magnetic field, parallel to the symmetry axis of the radiation field, the Hanle effect does not operate and the polarization signals in the core of the lines are unaffected when making the partial frequency redistribution treatment, under the angle-averaged assumption considered in this paper. The angle-dependent case will be studied in a future work. In the K line, the magnetic field significantly affects the wing polarization through magneto-optical effects, reducing $Q/I$ and enhancing $U/I$ signals (see Figure \ref{Fig_17_Vertical_B_mu_01_K_line}). In contrast, the linear polarization of the $8542 \: \AA$ line remains largely unaffected by such magneto-optical effects because the IR triplet is rather insensitive to the effects of PRD and J-state interference, with changes in its linear polarization arising only from the Zeeman effect at large magnetic field strengths (see Figure \ref{Fig_17_Vertical_B_mu_01_8542_A}).\\

Finally, comparison of the synthetic circular polarization profiles to those obtained applying the WFA shows a very good agreement for the lines of the IR triplet and in the core region of the H and K lines for the considered field strengths. However, the application of the WFA leads to an overestimation of the outer lobes of the circular polarization of the H and K lines. On the other hand, we find that carrying out the synthesis while neglecting atomic level polarization leads to an underestimation of the same lobes in $V$ (see Figure \ref{Fig_11_Mag_K_8542_V_pol_nopol}), while also correctly reproducing the profiles in the IR triplet.\\

Given the recent upgrade of the instrument GRIS installed at the GREGOR telescope of the Observatorio del Teide, and the high-quality data obtained by the SCIP and SUSI instruments aboard the Sunrise III balloon-borne telescope, we expect a large number of new investigations based on the Stokes profiles of the resonance and subordinate lines of Ca~{\sc{ii}}. We believe that the theoretical results we present in this paper, obtained through detailed radiative transfer calculations in 1D and static models of the solar chromosphere, will be very useful for understanding such spectropolarimetric observations. Given that the chromospheric plasma is highly inhomogeneous and dynamic, we plan to pursue further theoretical investigations using more realistic chromospheric models and partial redistribution treatments.\\

\begin{acknowledgements}
We acknowledge support from the Agencia Estatal de Investigaci\'on del Ministerio de Ciencia, Innovaci\'on y Universidades (MCIU/AEI) under the grant ``Polarimetric Inference of  Magnetic Fields'' and the European Regional Development Fund (ERDF) with reference PID2022-136563NB-I00/10.13039/501100011033. E.A.B.’s participation was funded by the European Research Council through Synergy grant No. 810218 (“The Whole Sun” ERC-2018-SyG). T.P.A.'s participation in the publication is part of the Project RYC2021-034006-I, funded by MICIN/AEI/10.13039/501100011033, and the European Union “NextGenerationEU”/RTRP.
\end{acknowledgements}

\bibliographystyle{aa}
\bibliography{bibliography}

\appendix
\clearpage

\section{The impact of the Zeeman effect on the linear polarization of the K line} \label{Appendix_forward_scatt_trans_Zeeman}
    
    \begin{strip}
      \centering
      \begin{minipage}[t]{0.48\textwidth}
        The linear polarization profile of the K line, when modeled considering PRD in a magnetized atmosphere, exhibits side peaks in either side of the center of the line, at $\Delta \lambda \sim 0.15 \: \AA$, as shown in Figure \ref{Fig_10_Mag_K_line_diff_B_FALC} for the forward scattering geometry. In order to analyze the physical origin of these peaks, we performed spectral syntheses in the same setup as in Section \ref{Subsection_Mag_mu1}, but neglecting atomic polarization and ignoring the axial symmetry breaking and anisotropy of the radiation field (hereafter the unpolarized case). This configuration has already been described in Section \ref{Section_circular_pol}, but we analyze the $Q/I$ linear polarization profiles in the present Appendix. \\
    
        Figure \ref{Fig_13_Mag_K_8542_Q_pol_nopol} shows the linear polarization profiles $Q/I$ for the standard and the unpolarized case in the K and $8542 \: \AA$ lines, computed with magnetic fields of different strength. In the unpolarized case, linear polarization signals can only be produced via the Zeeman effect. For $10 \: \mathrm{G}$, we see no $Q/I$ signal in the unpolarized case, so the Zeeman effect makes no contribution. Thus, the side peaks at $\Delta \lambda \sim 0.15 \: \AA$ in the K line appear as a result of the combined effect of the PRD treatment and the Hanle effect.\\
         
      \end{minipage}
      \hfill
      \begin{minipage}[t]{0.48\textwidth}
        In the $200 \: \mathrm{G}$ case, the polarization profile of the $8542 \: \AA$ line is almost completely dominated by the Zeeman effect; the same occurs for the H line and the rest of the IR triplet. However, the core region of the K line is still governed by scattering polarization and its modification by the Hanle effect. Actually, the amplitude of the signal in the core region is very similar for both field strengths, as the Hanle effect of the upper level is already approaching saturation for $10 \: \mathrm{G}$. The Zeeman effect has a clear contribution to the side peaks of the K line in the $200 \: \mathrm{G}$ case, but the amplitude difference between the standard and the unpolarized case is still significant. In fact, the difference in amplitude between the polarization peaks  produced only due to the Zeeman effect (unpolarized case) and in the standard full-physics case (accounting for atomic polarization) is the same for any magnetic field near or over the Hanle saturation value of the upper level. We conclude that the combined action of the PRD and the Hanle effects produces side peaks in the linear polarization profile even in the absence of transverse magnetic fields strong enough to produce appreciable Zeeman signals.\\ 
      \end{minipage}

    \vspace{1.5cm}
      \includegraphics[width=0.8\textwidth]{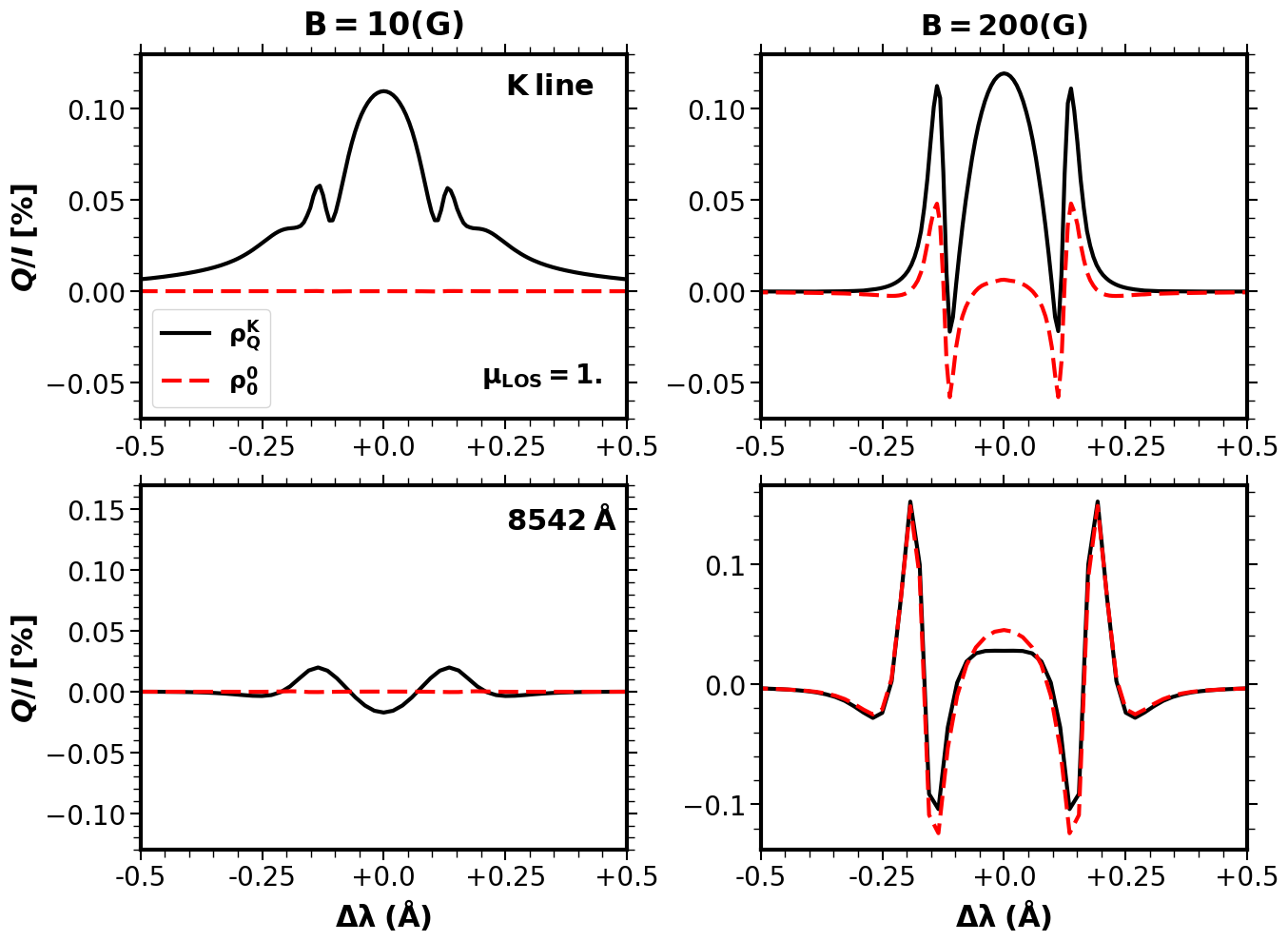}
        \captionof{figure}{Fractional linear polarization profiles $Q/I$ in forward scattering ($\mu = 1$) as a function of wavelength distance from the center of each line. The standard synthesis (computed with full physical considerations) is represented in solid black (dubbed $\rho^{K}_{Q}$) and the unpolarized case in dashed red ($\rho^{0}_{0}$). We show results corresponding to syntheses with $10 \: \mathrm{G}$ (left panels) and $200 \: \mathrm{G}$ (right panels) horizontal magnetic fields. The K and $8542 \: \AA$ lines are shown in the upper and lower panels, respectively.}
        \label{Fig_13_Mag_K_8542_Q_pol_nopol}
    \end{strip} 

\clearpage
\twocolumn

\section{Stokes profiles for an atmospheric model representative of a solar plage (FAL-P)} \label{Appendix_FAL_P}
    \begin{strip}
      \centering
      \begin{minipage}[t]{0.48\textwidth}
        In this Appendix, we present polarization profiles computed for the atmospheric model P of \cite{FAL_C}, representative of a solar plage region. The qualitative results presented throughout this work, concerning the redistribution treatment and the multi-level and multi-term atomic models, hold for this atmosphere. Figures \ref{Fig_6_Mag_K_line_diff_B_FALP} and \ref{Fig_5_Mag_IRT_diff_B_FALP} are equivalent to Figures \ref{Fig_6_Mag_K_line_diff_B_FALC} and \ref{Fig_5_Mag_IRT_diff_B_FALC}, but considering model P instead of C.\\

        Regarding the magnetic sensitivity, the qualitative behavior of the H and K lines remains consistent with the results presented throughout this study, both for the line-core region (see Figure \ref{Fig_6_Mag_K_line_diff_B_FALP}) and the inference pattern between the lines. However, for the IR triplet lines (see Figure \ref{Fig_5_Mag_IRT_diff_B_FALP}) the depolarization observed in the $Q/I$ signal due to sub-gauss magnetic fields is significantly weaker than in the FAL-C case. \\

        The modification produced by the magnetic field in the polarization of the $8542 \: \AA$ and $8662 \: \AA$ IR triplet lines is dominated by the lower level Hanle effect \citep{MSTB_2010}. As shown by the right-hand side of Equation \ref{Eq_Hanle_parameter}, the\\
      \end{minipage}
      \hfill
      \begin{minipage}[t]{0.48\textwidth}
         efficiency in depolarization is inversely proportional to the lifetime of the metastable levels, which is determined by the absorption process. Consequently, a stronger radiation field (corresponding to a larger value of $J_{ul}$) implies a larger critical magnetic field strength for the onset of the Hanle effect in the lower levels of the IR triplet lines. Around the formation height of these lines, the radiation field is stronger when considering the FAL-P atmospheric model compared to FAL-C, implying that the polarization profiles for the former atmospheric model are sensitive to stronger magnetic fields.\\

        On the other hand, for magnetic fields in the gauss range, the magnetic sensitivity of the K line is governed by the upper level Hanle effect, where the lifetime of the level is determined by the Einstein coefficient for spontaneous emission (see left-hand side expression in Equation \ref{Eq_Hanle_parameter}). The spontaneous emission lifetime of a given level is independent of the radiation field, so in the gauss range the magnetic sensitivity of the K line is independent of the atmospheric model.\\
      \end{minipage}

    \vspace{1.5cm}

    \includegraphics[width=0.85\textwidth]{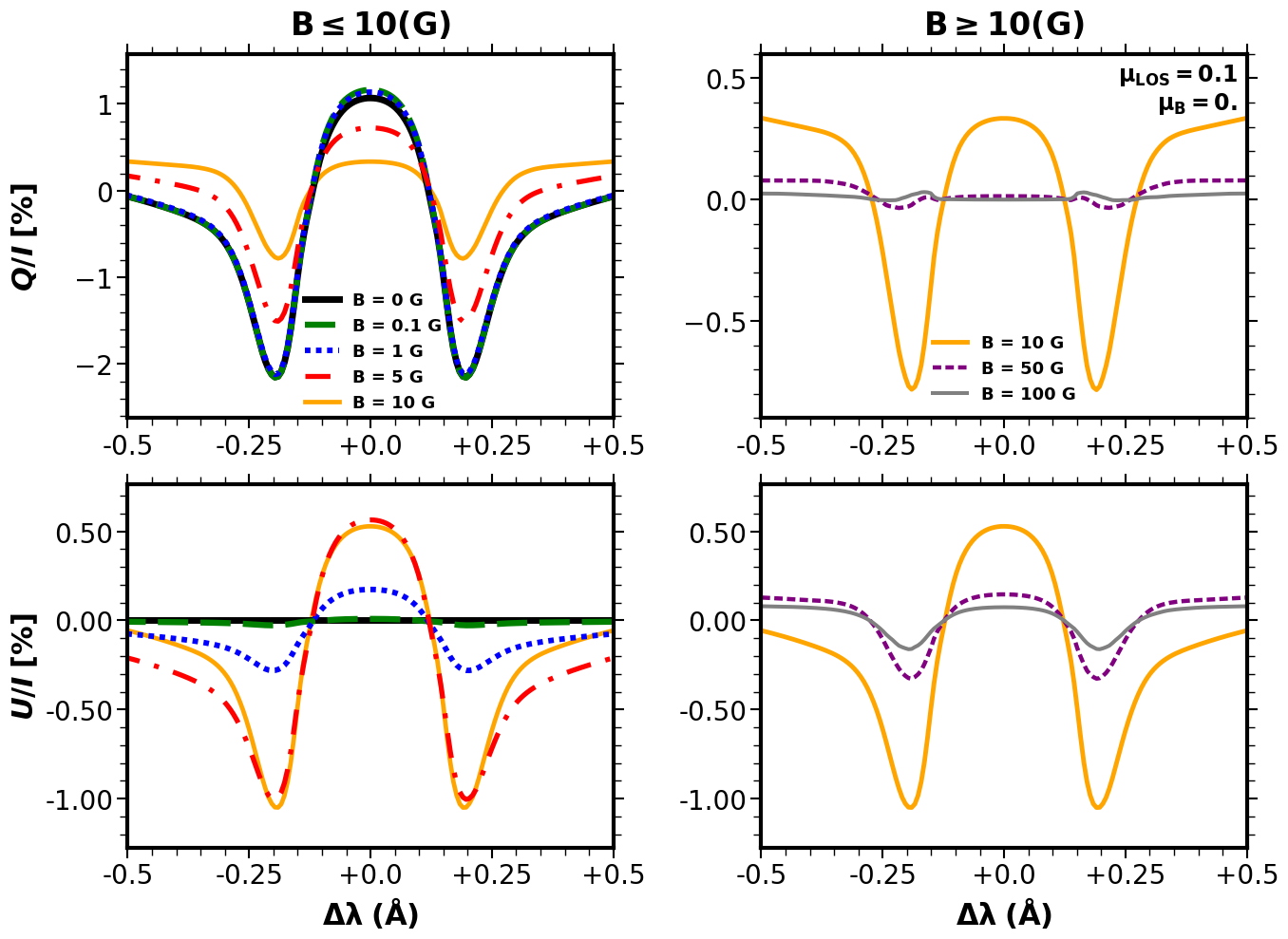}
    \captionof{figure}{Same as Figure \ref{Fig_6_Mag_K_line_diff_B_FALC}, but for the plage-like semiempirical atmospheric model FAL-P.}
    \label{Fig_6_Mag_K_line_diff_B_FALP}

    \vspace{.5cm}
    
    \includegraphics[width=1.\textwidth]{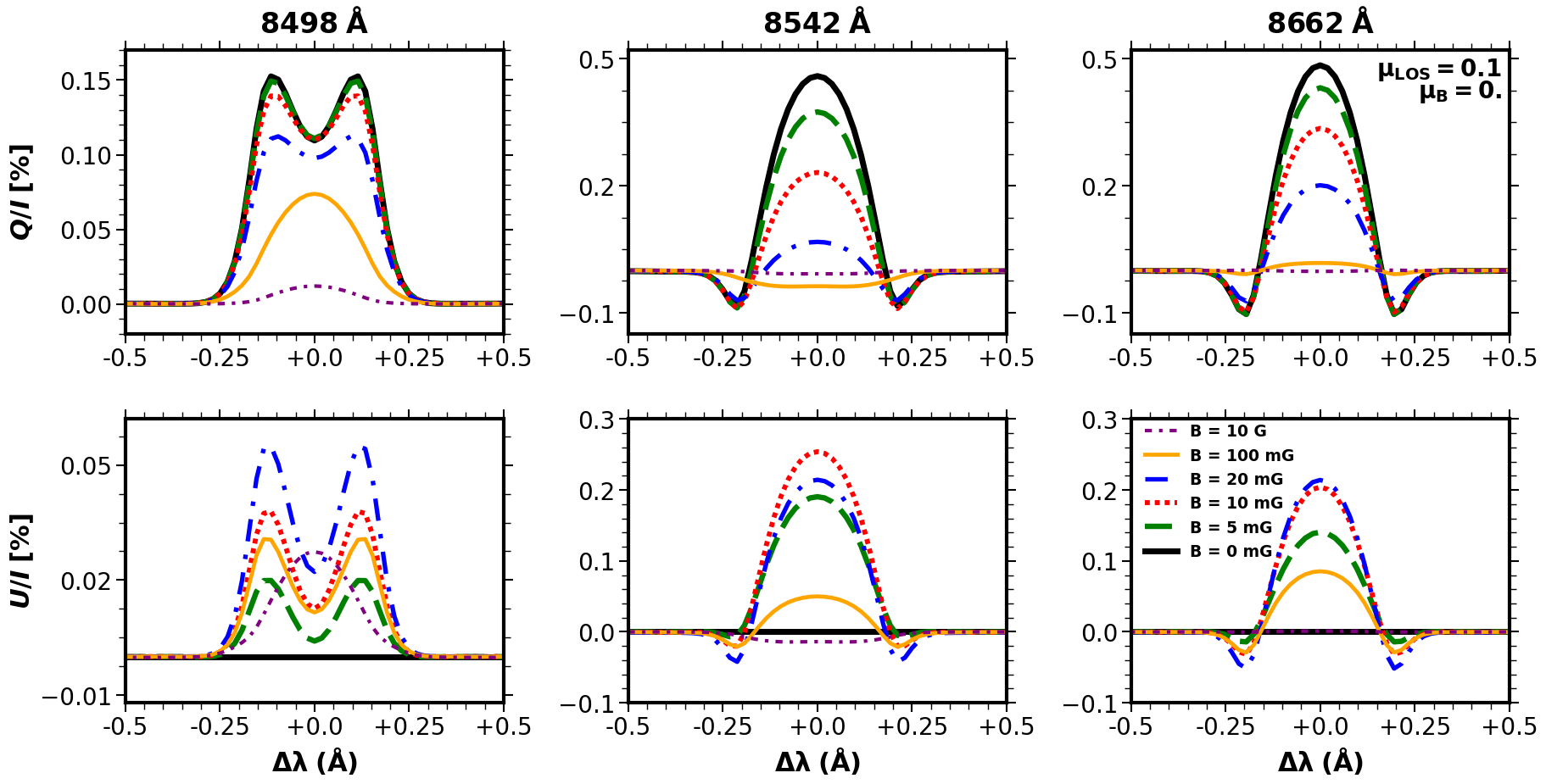}
    \captionof{figure}{Same as Figure \ref{Fig_5_Mag_IRT_diff_B_FALC}, but for the plage-like semiempirical atmospheric model FAL-P.}
    \label{Fig_5_Mag_IRT_diff_B_FALP}
        
    \end{strip}

\section{Implemented physics} \label{Appendix_collisions}
    
    In this Appendix, we present information on the specific physics implemented in the syntheses. Although the atmospheric and atomic models and the frequency redistribution treatments have already been presented in Section \ref{Section_Formulation}, there are several technical details and physical treatments that need to be explained in order to facilitate the reproducibility of our results.
    
    \subsection*{Technical details}
    The spectral synthesis module of the HanleRT-TIC radiative transfer code allows to compute the intensity and polarization solutions concurrently; this is, the $\rho^{0}_{0}$ population multipole components of the atomic levels are not fixed when the atomic polarization multipoles ($\rho^{K}_{Q}$ with $K > 0$) are computed. The intensity problem is solved first, but the atomic population multipole components ($\rho^{0}_{0}$) are not fixed when solving the full Stokes RT problem. This allows for ionization processes to be considered in the calculations of the atomic level populations. Nevertheless, we checked various cases where we implemented a two step solution (computing $\rho^{0}_{0}$ first and calculating $\rho^{K}_{Q}$ after fixing the populations of all levels in the atomic model) and the results did not change.\\
    
    The angular quadrature is composed of 8 polar nodes per hemisphere and 2 axial nodes per octant. The convergence criteria to determine that the self-consistent solution has been reached is set at a maximum relative change (MRC) of $10^{-4}$ for the $\rho^{0}_{0}$ multipoles and $10^{-3}$ for the other $\rho^{K}_{Q}$ multipoles. The values of the resolution parameters for the frequency redistribution integral are specified in table \ref{Table_Freq_int_resol}, and the meaning of each one can be consulted in the documentation of the HanleRT-TIC code. Leaving these parameter values as predetermined in the code can lead to small numerical asymmetries in the polarization profiles in the core of the K line whenever PRD is implemented, so we used the values in table \ref{Table_Freq_int_resol} for a correct integration.\\
    
    \subsection*{Physical models and treatment}
    We include the following atoms as background contributions, but explicitly without spectral lines: He, C, N, O, Na, Al, Si, S, Fe, Ni, Mg. Moreover, the molecules included as background contributions and regarding chemical equilibrium are: $\mathrm{C_{2}}$, CH, CN, CO, $\mathrm{H_{2}}$, $\mathrm{H^{+}_{2}}$, $\mathrm{H_{2}O}$, $\mathrm{N_{2}}$, NH, NO, $\mathrm{O_{2}}$, OH. The atomic models for each one can be found in the HanleRT-TIC repository.\\
    
    Regarding the Ca~{\sc{ii}} atomic model, the energy levels and the Einstein coefficients for spontaneous emission are taken from the NIST database \citep{NIST}. These values can suffer small changes as more advanced experiments are carried out, but the implemented values are those publicly available at the date of this work. For the multi-term atomic model, the spontaneous emission coefficients between terms are computed with the following relation:
    
    \begin{equation}
        \quad \quad A(L_{u} \rightarrow L_{l}) = \frac{\sum_{J_{u}} (2J_{u}+1) \sum_{J_{l}}A(L_{u}, J_{u} \rightarrow L_{l}, J_{l})}{\sum_{J_{u}} (2J_{u}+1)} \; ,
    \end{equation}
    
    \noindent where $A(L_{u}, J_{u} \rightarrow L_{l}, J_{l})$ are the already mentioned Einstein coefficients for the multi-level atom. \\
    
    The opacity fudge parameter is the same as in the RH code \citep{Uitenbroek_2001}. The inelastic collisions, which are defined as those that change the atomic level populations and are produced by collisions with free electrons, are implemented following \citet{Shine1974}. The $C_{ul} / N_{e}$ bound-bound collisional rate values are presented in Table 1 of \citet{MSTB_2010}, with $N_{e}$ being the electron density. The bound-free rates are extracted from the RH code \citep{Uitenbroek_2001} and converted to $\mathrm{cm^{3}s^{-1}}$ units. The forbidden collisional transitions can transfer atomic polarization among atomic levels. These transfer rates are quantified by the $\mathrm{C^{(K)}}$ multipole components with $\mathrm{K > 0}$, which we compute from $C^{(0)}$ following Appendix 4 of \citealt{Landi_2004}. Indeed, this has a significant impact in the polarization of the $8542 \: \AA$ and $8662 \: \AA$ lines, as the amplitude of the linear polarization in the unmagnetized case is $\sim 0.3\%$ lower when forbidden collisional transitions are not implemented.\\
    
    The elastic collisions with neutral hydrogen atoms, which reduce the atomic polarization of a given level, are implemented following \citet{MS_2014}. Here, the total depolarization rate of an atomic level is defined as 
    
    \begin{equation*}
       \quad \quad  \quad \quad  \quad \quad g^{k}(\alpha J) = \bar{C}(\alpha J) + D^{K}(\alpha J) \; ,
    \end{equation*}
    
    \noindent where $D^{K}(\alpha J)$ are the elastic depolarization rates defined in equation (7.102) of \citet{Landi_2004}. This collisions are also named ``quasi-elastic'' because $\bar{C}(\alpha J)$ is
    
    \begin{equation*}
        \quad \quad  \quad \quad  \quad \quad \bar{C}(\alpha J) = \sum_{\alpha' J'} C^{(0)}(\alpha' J' \leftarrow \alpha J) \; ,
    \end{equation*}
    
    \noindent where the sum over $\alpha' J'$ is restricted to atomic levels in the same term as $\alpha J$. $\bar{C}(\alpha J)$ describes the rate of collisional transitions between atomic levels pertaining to the same term and not carrying atomic polarization. All in all, the following expressions are implemented in the SE equations:
    
    \begin{align*}
         \quad \quad  \quad \quad  \quad \quad & \bar{C} = a_{0} 10^{-9} \left(\frac{T}{5000}\right)^{b_{0}} N_{H} \\ &g^{k} = a_{k} 10^{-9} \left(\frac{T}{5000}\right)^{b_{K}} N_{H} \quad \mathrm{with} \quad K > 0 \; ,
    \end{align*}
    
    \noindent where $T$ and $N_{H}$ are the temperature and neutral hydrogen density, respectively. The values for $a_{K}$ and $b_{K}$ are tabulated for Ca~{\sc{ii}} in Table 1 of \citet{MS_2014}. \\

    These depolarizing collisional rates describe the relaxation of the density matrix multipoles in a multi-level atomic model. To our knowledge, an equivalent formalism has not yet been developed for multi-term atoms. In such models, the density matrix includes terms that account for quantum interference between different $J$ levels belonging to the same atomic term. Collisions with neutral hydrogen atoms are expected to affect this interference as well, reducing it at increasing neutral hydrogen number densities. Since no theoretical or numerical treatment of these rates is available, and setting them to zero would be inconsistent, we have adopted a weighted average (by statistical weight) of the depolarizing collision rates of the involved $J$ levels, namely
    
    \begin{equation*}
       \quad \quad  \quad \quad  \quad   D^{K}(J,J') = \frac{(2J+1)D^{K}(J) + (2J'+1)D^{K}(J')}{2J+1+2J'+1} \; .
    \end{equation*}\\

    \begin{table}
        \caption{Values for the frequency redistribution integral parameters for Ca~{\sc{ii}} modeling in HanleRT-TIC.}
        \label{Table_Freq_int_resol}
        \small
        \centering
        \begin{tabular}{c c c}
        \hline
        RED\_RESO & RED\_NEGL & RED\_CORE \\
        3.5 & $10^{4}$ & 2.5 \\
        \hline
        RED\_RANG & RED\_RANG\_CORE & RED\_VLAR \\
        3.5 & 3.5 & 7 \\
        \hline
        RED\_VLAR\_CORE & RED\_FSTP & RED\_FSTP\_CORE \\
        7 & 0.5 & 0.15 \\
        \hline
        RED\_MSTP & RED\_MSTP\_CORE & \\
        4 & 13.33 & \\
        \hline
        \end{tabular}        
    \end{table}

\end{document}